\documentclass[11pt,letter]{emulateapj}
\usepackage{amsmath}
\usepackage{graphicx}
\usepackage{epstopdf}
\usepackage[T1]{fontenc}
\PassOptionsToPackage{hyphens}{url}\usepackage{hyperref}

\def\alf{Alfv\'en}
\newcommand{\QU}{$Q$-$U$\,}

\shorttitle{Applied Polarization Diagnostics}
\shortauthors{Herron et al.}
\begin{document}

\title{Advanced Diagnostics for the Study of Linearly Polarized Emission. II: Application to Diffuse Interstellar Radio Synchrotron Emission}

\author{C.~A.~Herron\altaffilmark{1}, Blakesley Burkhart\altaffilmark{2}, B.~M.~Gaensler\altaffilmark{3,1}, G.~F.~Lewis\altaffilmark{1}, N.~M.~McClure-Griffiths\altaffilmark{4}, G.~Bernardi\altaffilmark{5,6}, E.~Carretti\altaffilmark{7}, M.~Haverkorn\altaffilmark{8}, M.~Kesteven\altaffilmark{9}, S.~Poppi\altaffilmark{7}, L.~Staveley-Smith\altaffilmark{10,11}}

\altaffiltext{1}{Sydney Institute for Astronomy, School of Physics, A28, The University
of Sydney, NSW, 2006, Australia; C.Herron@physics.usyd.edu.au}

\altaffiltext{2}{Harvard-Smithsonian Center for Astrophysics, 60 Garden St. Cambridge, MA, USA}

\altaffiltext{3}{Dunlap Institute for Astronomy and Astrophysics, University of Toronto, 50 St. George Street,
Toronto, Ontario, M5S 3H4, Canada}

\altaffiltext{4}{Research School of Astronomy and Astrophysics, The Australian National University, Canberra, ACT 2611, Australia}

\altaffiltext{5}{Square Kilometre Array South Africa (SKA SA), Park Road, Pinelands 7405, South Africa}

\altaffiltext{6}{Department of Physics and Electronics, Rhodes University, P.O. Box 94, Grahamstown, 6140, South Africa}

\altaffiltext{7}{INAF Osservatorio Astronomico di Cagliari, Via della Scienza 5, 09047 Selargius (CA), Italy}

\altaffiltext{8}{Department of Astrophysics/IMAPP, Radboud University Nijmegen, PO Box 9010, 6500 GL Nijmegen, the Netherlands}

\altaffiltext{9}{CSIRO Astronomy and Space Science, PO Box 76, Epping, NSW 1710, Australia}

\altaffiltext{10}{International Centre for Radio Astronomy Research, University of Western Australia, Crawley, WA 6009, Australia}

\altaffiltext{11}{ARC Centre of Excellence for All-sky Astrophysics (CAASTRO)}

\date{\today}


\begin{abstract}

Diagnostics of polarized emission provide us with valuable information on the Galactic magnetic field and the state of turbulence in the interstellar medium, which cannot be obtained from synchrotron intensity alone. In Paper I \citep{Herron2017c}, we derived polarization diagnostics that are rotationally and translationally invariant in the \QU plane, similar to the polarization gradient. In this paper, we apply these diagnostics to simulations of ideal magnetohydrodynamic turbulence that have a range of sonic and Alfv\'enic Mach numbers. We generate synthetic images of Stokes $Q$ and $U$ for these simulations, for the cases where the turbulence is illuminated from behind by uniform polarized emission, and where the polarized emission originates from within the turbulent volume. From these simulated images we calculate the polarization diagnostics derived in Paper I, for different lines of sight relative to the mean magnetic field, and for a range of frequencies. For all of our simulations, we find that the polarization gradient is very similar to the generalized polarization gradient, and that both trace spatial variations in the magnetoionic medium for the case where emission originates within the turbulent volume, provided that the medium is not supersonic. We propose a method for distinguishing the cases of emission coming from behind or within a turbulent, Faraday rotating medium, and a method to partly map the rotation measure of the observed region. We also speculate on statistics of these diagnostics that may allow us to constrain the physical properties of an observed turbulent region. 

\end{abstract}
\section{Introduction} 
\label{intro} 

Turbulence and magnetic fields are both ubiquitous throughout the multi-phase interstellar medium (ISM) (see \citealt{Armstrong1995} and \citealt{Haverkorn2015a} respectively), and have a large impact on the formation of stars (e.g. \citealt{Elmegreen2004, Scalo2004, McKee2007, FalcetaGoncalves2014, Hennebelle2014, Federrath2015}), the exchange of gas between the disk and the halo of the Milky Way (e.g. \citealt{Joung2012, Beck2013}), and the stellar life cycle as a whole \citep{Ferriere2001}. 

Whereas the formation of stars from turbulent molecular clouds is a focal point for current research, less emphasis is placed on the diffuse warm-ionized medium. As the turbulence in the cold-neutral medium is inherited from the warm-ionized medium (see the review by \citealt{McKee2007}, and references therein), a greater understanding of the properties of the turbulence in the warm-ionized medium will provide us with an enhanced understanding of the lifecycle of interstellar gas. Additionally, the warm-ionized medium provides us with unique probes of the interstellar magnetic field \citep{Haverkorn2013}, which can be used to study the structure and evolution of the Galactic magnetic field \citep{Beck2013, Haverkorn2015a}, with implications for the history of star formation in the Milky Way. 

The diffuse warm-ionized medium can be studied by observing H$\alpha$ emission (e.g. the Wisconsin H Alpha Mapper, \citealt{Haffner2003}), or it can be studied at radio wavelengths by observing the linearly polarized synchrotron emission radiated by ultra-relativistic electrons, that are spiralling around magnetic field lines \citep{Ginzburg1965}. Recently it was found that the statistics of total synchrotron intensity can provide us with information on the orientation of the mean magnetic field relative to the line of sight \citep{Lazarian2017}, the compressibility of the magnetoionic medium (any magnetized and ionized medium,  \citealt{Lazarian2012}), and how the turbulence is being driven \citep{Herron2017a}. \cite{Herron2016} investigated whether a statistical analysis of mock synchrotron intensity images could be used to constrain properties of the turbulence, such as the sonic and Alfv\'enic Mach numbers, given by
\begin{equation}
\mathcal{M}_s = \left< \frac{|\boldsymbol{v}|}{c_s} \right>, \quad \text{and} \quad \mathcal{M}_A = \left< \frac{|\boldsymbol{v}|}{v_A} \right> , \label{Mach_nums}
\end{equation}
respectively, where $|\boldsymbol{v}|$ is the amplitude of the velocity vector $\boldsymbol{v}$, $c_s$ is the sound speed, and $v_A$ = $|\boldsymbol{B}|/\sqrt{\rho}$ is the Alfv\'en velocity, calculated from the amplitude of the magnetic field $\boldsymbol{B}$, and the density $\rho$. We use angled brackets to denote an average over the turbulent volume. \cite{Herron2016} found that statistics of synchrotron intensity are sensitive to the Alfv\'enic Mach number, however they concluded that additional constraints are required to determine the Mach numbers, which could be provided by statistics of polarization diagnostics.

Polarization diagnostics that are rotationally and translationally invariant in the Stokes \QU plane, such as the spatial polarization gradient \citep{Gaensler2011, Burkhart2012}, have great potential to provide robust statistics that we can use to constrain the regime of turbulence, as they are unaffected by the limitations of interferometric data, such as missing interferometer spacings. The polarization gradient is given by \citep{Gaensler2011}
\begin{equation}
{|\nabla {\boldsymbol{P}}|}={\sqrt{\biggl(\frac{\partial Q}{\partial x}\biggr)^2
+\biggl(\frac{\partial U}{\partial x}\biggr)^2
+\biggl(\frac{\partial Q}{\partial y}\biggr)^2
+\biggl(\frac{\partial U}{\partial y}\biggr)^2},}
\label{p_grad}
\end{equation}
where $x$ and $y$ are the horizontal and vertical axes of the image plane, respectively, and $\boldsymbol{P} = Q + iU$ is the complex polarization. \cite{Gaensler2011} found that the polarization gradient traces spatial variations in the magnetoionic medium, and \cite{Burkhart2012} found that statistics of the polarization gradient, such as the skewness and genus, were sensitive to the sonic Mach number of their simulations. However, \cite{Herron2017b} cast doubt on the ability of the skewness of the polarization gradient to probe the regime of turbulence, as they found that the skewness of the gradient was very sensitive to angular resolution, and the size of the evaluation box used to calculate the skewness, in the Canadian Galactic Plane Survey dataset (CGPS, \citealt{Landecker2010}). 

Statistics of polarized emission that provide insight on the properties of an observed turbulent region have also been developed by \cite{Lazarian2016}, which involve correlation functions of the polarized emission. These statistics have been applied to simulated turbulence by \cite{Lee2016} and \cite{Zhang2016}, and to optical observations of blazar emission by \cite{Guo2017}.

\begin{widetext}
In \citet[hereafter Paper I]{Herron2017c} we derived new polarization diagnostics that are rotationally and translationally invariant in the \QU plane, to work towards the discovery of complementary methods of constraining properties of turbulence. These diagnostics include the:
\begin{itemize}
\item Generalized polarization gradient - Traces spatial changes in the observed complex polarization, and reduces to the polarization gradient in the case of uniform polarized emission illuminating a turbulent region from behind. Like the polarization gradient, this quantity may trace vorticity, shear, or shocks in the turbulence. Given by Eq. \ref{eq:direc_max}, where $s$ denotes distance in the image plane.
\begin{align}
\bigg| \frac{\partial \boldsymbol{P}}{\partial s} \bigg|_{\text{max}} &= \Biggl[ \frac{1}{2} \biggl( \biggl(\frac{\partial Q}{\partial x} \biggr)^2 + \biggl(\frac{\partial U}{\partial x} \biggr)^2 + \biggl(\frac{\partial Q}{\partial y} \biggr)^2 + \biggl(\frac{\partial U}{\partial y} \biggr)^2  \biggr) + \nonumber \\ 
 & \frac{1}{2} \sqrt{\biggl( \biggl(\frac{\partial Q}{\partial x} \biggr)^2 + \biggl(\frac{\partial U}{\partial x} \biggr)^2 + \biggl(\frac{\partial Q}{\partial y} \biggr)^2 + \biggl(\frac{\partial U}{\partial y} \biggr)^2 \biggr)^2 - 4 \biggl( \frac{\partial Q}{\partial x} \frac{\partial U}{\partial y} - \frac{\partial Q}{\partial y} \frac{\partial U}{\partial x} \biggr)^2  } \Biggr]^{1/2}.  \label{eq:direc_max}
\end{align}

\item Radial and tangential components of the polarization directional derivative - Trace how changes in polarization intensity and polarization angle respectively contribute to the polarization directional derivative (although these are not invariant). May provide insight on whether small-scale or large-scale spatial variations in the turbulence are primarily responsible for the observed polarization. The maximum value of the radial component is given by
\begin{equation}
\frac{\partial \boldsymbol{P}}{\partial s}_{\text{rad, max}} = \sqrt{ \frac{ \bigl( Q \frac{\partial Q}{\partial x} + U \frac{\partial U}{\partial x} \bigr)^2 + \bigl( Q \frac{\partial Q}{\partial y} + U \frac{\partial U}{\partial y} \bigr)^2 }{ Q^2 + U^2 } }, \label{rad_direc_max}
\end{equation}
and the maximum value of the tangential component is given by
\begin{equation}
\frac{\partial \boldsymbol{P}}{\partial s}_{\text{tang, max}} = \sqrt{ \frac{ \bigl( Q \frac{\partial U}{\partial x} - U \frac{\partial Q}{\partial x} \bigr)^2 + \bigl( Q \frac{\partial U}{\partial y} - U \frac{\partial Q}{\partial y} \bigr)^2 }{ Q^2 + U^2 } }. \label{tang_direc_max}
\end{equation}

\item Polarization directional curvature - Traces second order spatial changes in the observed polarization, and is independent of the generalized polarization gradient, so may provide a new way of visualizing turbulence. At a wavelength $\lambda$, it is given by
\begin{align}
k_s(x,y,\lambda^2;\theta) &=  \bigg| \frac{\partial \boldsymbol{P}}{\partial s} \bigg|^{-3} \biggl[ \cos^3 \theta \biggl( \frac{\partial Q}{\partial x} \frac{\partial^2 U}{\partial x^2} - \frac{\partial U}{\partial x} \frac{\partial^2 Q}{\partial x^2} \biggr) + 2 \cos^2 \theta \sin \theta \biggl( \frac{\partial Q}{\partial x} \frac{\partial^2 U}{\partial x \partial y} - \frac{\partial U}{\partial x} \frac{\partial^2 Q}{\partial x \partial y} \biggr) + \nonumber \\
&  \cos^2 \theta \sin \theta \biggl( \frac{\partial Q}{\partial y} \frac{\partial^2 U}{\partial x^2} - \frac{\partial U}{\partial y} \frac{\partial^2 Q}{\partial x^2} \biggr) + 2 \cos \theta \sin^2 \theta \biggl( \frac{\partial Q}{\partial y} \frac{\partial^2 U}{\partial x \partial y} - \frac{\partial U}{\partial y} \frac{\partial^2 Q}{\partial x \partial y} \biggr) + \nonumber \\
&  \cos \theta \sin^2 \theta \biggl( \frac{\partial Q}{\partial x} \frac{\partial^2 U}{\partial y^2} - \frac{\partial U}{\partial x} \frac{\partial^2 Q}{\partial y^2} \biggr) + \sin^3 \theta \biggl( \frac{\partial Q}{\partial y} \frac{\partial^2 U}{\partial y^2} - \frac{\partial U}{\partial y} \frac{\partial^2 Q}{\partial y^2} \biggr) \biggr].\label{eq:pol_direc_curv}
\end{align}

\item Polarization wavelength derivative - Traces spectral changes in the observed polarization at a pixel of an image. May provide new insight on turbulent Faraday rotation. Given by
\begin{equation}
\bigg| \frac{\partial \boldsymbol{P}}{\partial \lambda^2} \bigg| = \sqrt{\biggl( \frac{\partial Q}{\partial \lambda^2} \biggr)^2 + \biggl( \frac{\partial U}{\partial \lambda^2} \biggr)^2}. \label{pol_wav_deriv}
\end{equation}

\item Polarization wavelength curvature - Traces second order spectral changes in the observed polarization at a pixel of an image. Together with the polarization wavelength derivative, these diagnostics may provide a new robust method of studying Faraday rotation. Given by
\begin{equation}
k_{\lambda}(x,y,\lambda^2) = \bigg| \frac{\partial \boldsymbol{P}}{\partial \lambda^2} \bigg|^{-3} \biggl[ \frac{\partial Q}{\partial \lambda^2} \frac{\partial^2 U}{\partial (\lambda^2)^2} - \frac{\partial U}{\partial \lambda^2} \frac{\partial^2 Q}{\partial (\lambda^2)^2} \biggr].
\end{equation} 

\item Polarization mixed derivative - Traces spatial and spectral changes in the observed polarization. Given by
\begin{align}
\bigg| \frac{\partial}{\partial \lambda^2} \biggl( \frac{\partial \boldsymbol{P}}{\partial s} \biggr) \bigg|_{\text{max}} &= \biggl[ \frac{1}{2} \biggl( \biggl( \frac{\partial^2 Q}{\partial \lambda^2 \partial x} \biggr)^2 + \biggl( \frac{\partial^2 U}{\partial \lambda^2 \partial x} \biggr)^2 + \biggl( \frac{\partial^2 Q}{\partial \lambda^2 \partial y} \biggr)^2 + \biggl( \frac{\partial^2 U}{\partial \lambda^2 \partial y} \biggr)^2 \biggr) + \nonumber \\
& \frac{1}{2} \sqrt{ \biggl( \biggl( \frac{\partial^2 Q}{\partial \lambda^2 \partial x} \biggr)^2 + \biggl( \frac{\partial^2 U}{\partial \lambda^2 \partial x} \biggr)^2 + \biggl( \frac{\partial^2 Q}{\partial \lambda^2 \partial y} \biggr)^2 + \biggl( \frac{\partial^2 U}{\partial \lambda^2 \partial y} \biggr)^2 \biggr)^2 - 4 \biggl( \frac{\partial^2 Q}{\partial \lambda^2 \partial x} \frac{\partial^2 U}{\partial \lambda^2 \partial y} - \frac{\partial^2 Q}{\partial \lambda^2 \partial y} \frac{\partial^2 U}{\partial \lambda^2 \partial x} \biggr)^2 } \biggr]^{1/2}. \label{mix_deriv_amp_max}
\end{align}
\end{itemize}
\end{widetext}

In this paper, we take a first step toward using these diagnostics to place robust statistical constraints on the physical properties of an observed turbulent region, by investigating the qualitative information about the observed turbulent region that is encoded in these diagnostics. To approach this problem we calculate mock images of Stokes $Q$ and $U$ for synchrotron emission radiated within or behind simulations of ideal magnetohydrodynamic (MHD) turbulence, and calculate the polarization diagnostics derived in Paper I from these images of $Q$ and $U$. We then compare the obtained diagnostics to physical properties of the turbulence, such as the rotation measure, for different lines of sight, observing frequencies, and for simulations in different regimes of turbulence.

We provide background information regarding polarized synchrotron emission and Faraday rotation in Section \ref{background}. In Section \ref{simulations} we describe our MHD simulations, and in Section \ref{synthetic} we describe the production of mock images of Stokes $Q$ and $U$ from the simulations. In Section \ref{internal_grad} we examine the polarization gradient and generalized polarization gradient, for the case where polarized emission is generated within the turbulent, Faraday-rotating volume. In Section \ref{rad_tang_comp}, we investigate how the radial and tangential components of the directional derivative can be used to compare the importance of large-scale and small-scale changes in the warm-ionized medium. In Section \ref{back_int} we discuss methods to distinguish between the cases where a turbulent medium is illuminated by background polarized emission, and where polarized emission comes from within the turbulent medium. In Section \ref{map_Fara} we outline a method to partly map the rotation measure of an observed turbulent region. In Section \ref{discuss} we discuss the qualitative information that can be gained from an analysis of the polarization diagnostics derived in Paper I, and speculate on what statistics will provide sensitive and robust probes of magnetoionic turbulence.

\section{Background} 
\label{background} 

To derive the intensity of synchrotron emission at a particular frequency, we need to consider the number density of ultra-relativistic electrons that radiate at this frequency. If we assume a homogeneous and isotropic power-law distribution in energy, $E$, then the number density $N(E)$ of ultra-relativistic electrons with energies between $E$ and $E + \mathrm{d}E$ is given by \citep{Ginzburg1965}
\begin{equation}
N(E) \, \mathrm{d} E = K E^{2 \alpha - 1} \, \mathrm{d} E, \label{energy_dist}
\end{equation}
for a normalization constant $K$, and spectral index $\alpha$, defined by intensity $I \propto \nu^{\alpha}$. The total intensity of the synchrotron emission at frequency $\nu$, $I(\nu)$ is then given by \citep{Ginzburg1965}
\begin{align}
I(\nu) &= \frac{e^3}{4 \pi m_e c^2} \int_0^{L} \frac{\sqrt{3}}{2 - 2 \alpha} \Gamma \left( \frac{2 - 6 \alpha}{12} \right) \Gamma \left( \frac{22 - 6 \alpha}{12} \right) \times \nonumber \\ 
& \quad \quad \left( \frac{3e}{2 \pi m_e^3 c^5} \right)^{-\alpha} K B_{\perp}^{1-\alpha} \nu^{\alpha} \, \mathrm{d} L', \label{sync_inten}
\end{align}
where $e$ is the charge of an electron, $m_e$ is the mass of an electron, $c$ is the speed of light, $L$ is the distance along the line of sight, over which we integrate the emissivity, $\Gamma$ is the gamma function, and $B_{\perp}$ is the strength of the magnetic field perpendicular to the line of sight. This radiation is linearly polarized, with linear polarization intensity $P$ determined from the fractional polarization $p$ according to \citep{Ginzburg1965}
\begin{equation}
p = \frac{P}{I} = \frac{3 - 3 \alpha}{5 - 3 \alpha}. \label{frac_pol}
\end{equation}
The plane of linear polarization is oriented to be perpendicular to the direction of the magnetic field projected onto the sky, and described by the polarization angle $\psi$, measured anti-clockwise from North (see \citealt{Gardner1966} and \citealt{Saikia1988} for more information on the polarization of synchrotron emission). The polarization intensity and polarization angle are related to the Stokes parameters $Q$ and $U$ according to $Q = P \cos 2 \psi$ and $U = P \sin 2 \psi$, or equivalently
\begin{equation}
P = \sqrt{Q^2 + U^2}, \text{ and } \psi = \frac{1}{2} \arctan \frac{U}{Q}.
\end{equation}
We can then define the complex polarization as $\boldsymbol{P} = Q + iU$, which is a vector in the complex \QU plane, whose modulus is $P$, and whose argument is $2\psi$.

Linearly polarized radio synchrotron emission possesses a unique ability to probe the turbulent magnetic field in the warm-ionized medium, because of the Faraday rotation the emission experiences as it propagates through a magnetoionic medium. If the intrinsic polarization angle of the synchrotron emission at the point it is emitted is $\psi_0$, then the observed polarization angle $\psi$ is given by
\begin{equation}
\psi = \psi_0 + \text{RM} \, \lambda^2, \label{Faraday_rotation}
\end{equation} 
where $\lambda$ is the wavelength of the emission, and $\text{RM}$ is the rotation measure, given by 
\begin{equation}
\text{RM} = 0.81 \int_{L}^{0} n_e B_{\parallel} \, \mathrm{d}z \, \, \text{rad m}^{-2}. \label{RM}
\end{equation}
In Eq. \ref{RM}, we define the $z$ axis to be along the line of sight, $n_e$ is the number density of electrons, measured in $\text{cm}^{-3}$, $B_{\parallel}$ is the strength of the magnetic field parallel to the line of sight, in $\mu G$, such that $B_{\parallel}$ is positive if the parallel component of the magnetic field is toward the observer, and we integrate from a position at $z=L$, measured in parsecs, toward the observer.

For the situation where we have a beam of polarized emission passing through a Faraday rotating volume, it is possible to determine the rotation measure by measuring how the polarization angle depends on $\lambda^2$, and hence it is possible to probe the electron density and magnetic field of the diffuse warm-ionized medium. In general, however, many sources of emission will be distributed along the line of sight, and the polarized emission from each source will experience a different rotation measure, causing the plane of polarization to rotate at a different rate. This causes the wavelength squared dependence of the observed polarization angle to be non-linear, and the rotation measure cannot be determined from fitting a linear slope to the dependence of the polarization angle on wavelength squared. 

The superposition of polarization vectors that have experienced differing degrees of Faraday rotation will also cause the observed polarization intensity to be lower than the scalar sum of the polarization intensity of each source, and this depolarization mechanism is referred to as differential Faraday rotation (see \citealt{Gardner1966} and \citealt{Sokoloff1998} for more information). Differential Faraday rotation is a form of depth depolarization, where emission is depolarized before reaching the observer, due to the superposition of polarization vectors with different polarization angles along the line of sight. Another form of depth depolarization is called wavelength independent depolarization, which occurs when the projection of the magnetic field onto the plane of the sky differs along the line of sight. This causes the intrinsic polarization angle along the line of sight to vary, and depolarization will still occur in the high frequency limit where Faraday rotation and differential Faraday rotation are negligible.

Depth depolarization mechanisms, which are sensitive to the turbulent fluctuations in the electron density and the magnetic field, complicate the link between the observed polarized emission and the magnetic field in the emitting region. This necessitates a statistical, wavelength-dependent approach to constraining the properties of observed magnetoionic turbulence using polarization diagnostics.

\section{Magnetohydrodynamic Simulations}
\label{simulations}

\begin{figure}
\begin{center}
\includegraphics[scale=0.32]{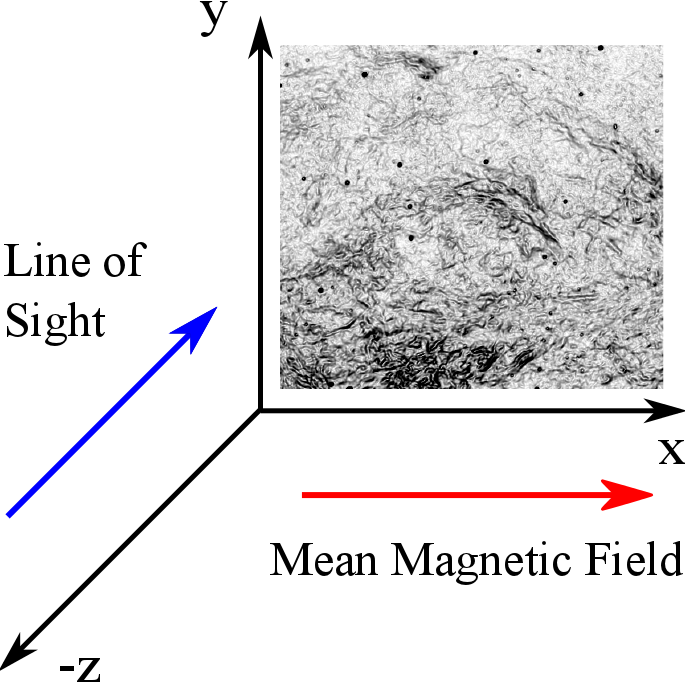}
\caption{A diagram illustrating how the simulations were set up. The mean magnetic field is in the $x$ direction, and we view the cube along the $x$, $y$, or $z$ directions, where the latter is shown above. The inset image shows polarization gradients from the Canadian Galactic Plane Survey (see \citealt{Herron2017b}), where black denotes a large amplitude of the polarization gradient, and white denotes a small amplitude.}
\label{sim_set_up}
\end{center}
\end{figure}

We use the same simulations of ideal MHD turbulence as those used by \cite{Gaensler2011}, \cite{Burkhart2012}, and \cite{Herron2016}. In this section, we will summarize the key properties of these simulations, and refer to \cite{Herron2016} for further details. The simulations are run using the second-order-accurate hybrid essentially non-oscillatory code produced by \cite{Cho2003}, which solves the ideal MHD equations with periodic boundary conditions. The simulations have $512$ pixels along each side, and all quantities are calculated in simulation units. Initially, each simulation cube has uniform pressure and density, and a uniform magnetic field oriented along the $x$ axis, as shown in Fig. \ref{sim_set_up}. The strength of the initial magnetic field can be altered to change the final Alfv\'enic Mach number of the simulation, and the initial pressure can be altered to change the final sonic Mach number. These simulations are driven solenoidally until the turbulence has sufficiently developed, assuming an isothermal equation of state, $p = c_s^2 \rho$, from which the sound speed can be calculated for each simulation. No assumptions were made regarding the components of the magnetic field parallel and perpendicular to the initial mean magnetic field.

The simulations that we analyze in this study are listed in Table \ref{TabSims}, which is reproduced from \cite{Herron2016}. Each simulation is assigned a code of the form Ms0.8Ma1.7, for example, which means that the simulation has a sonic Mach number of 0.8, and an Alfv\'enic Mach number of 1.7, in the temporal realisation of the simulation that is used in our analysis. As explained by \cite{Herron2016}, the Ms0.9Ma0.7 and Ms0.5Ma0.7 simulations are expected to be the simulations that best represent the warm-ionized medium of the Milky Way, as the sonic Mach number and average magnetic field strengths in these simulations are comparable to those measured in the Milky Way \citep{Hill2008, Sun2008, Gaensler2011, Iacobelli2014}.

For each simulation, we obtain dimensionless cubes of the thermal electron density, and each component of the magnetic and velocity field vectors. As we wish to calculate the Faraday rotation of polarized emission passing through these cubes using Eq. \ref{Faraday_rotation}, it is necessary to scale these dimensionless cubes to physical units. Following \cite{Burkhart2012}, we set the width of each pixel to be $0.15$ pc, so that the total width of each cube is $76.8$ pc. This is smaller than the scale height of the warm-ionized medium \citep{Gaensler2008}, and causes the driving scale of the simulations to be within the range of measured values for the outer scale on which turbulence in the warm-ionized medium is driven \citep{Haverkorn2008}, and thus should be a reasonable value. We set the average electron number density $\left< n_e \right>_0 = 0.2 \text{cm}^{-3}$, to equal the average electron density of the warm-ionized medium \citep{Ferriere2001, Haverkorn2013}. The mass density scaling $\rho_0$ is calculated from $\left< n_e \right>_0 $ by multiplying by the mass of a hydrogen atom. To define the velocity scaling, we use the same method as \cite{Hill2008}, so that our velocity scaling is given by $v_0 = 10.15 / \sqrt{p_{\text{ini}}}$, where $p_{\text{ini}}$ is the initial pressure in the simulation, in simulation units. We also use the same method as \cite{Hill2008} to define the scaling for the magnetic field, which is given by $B_0 = \sqrt{\rho_o v_0^2}$. 

We note that a consequence of this scaling is that supersonic simulations can have very large magnetic fields, because a large magnetic field is required to make the Alfv\'en speed similar to the high flow speed (in SI units, rather than simulation units) of these simulations.

\begin{table*}
\centering
\caption{The sonic and Alfv\'enic Mach numbers of each simulation used in this study, and the initial parameters used to run the simulation. Based on Table $1$ of \cite{Herron2016}.} \label{TabSims}
\begin{tabular}{| c c | c c | c c | c |}
\hline
\hline
Sim No. & Code & Init B (sim units) & Init P (sim units) & $M_s$ & $M_A$ & Turbulence Regime\\
\hline
1 & Ms11.0Ma1.4  & $0.1$ & $0.0049$ & $11.0$ & $1.4$ & Supersonic and super-\alf ic \\
2 & Ms9.2Ma1.8  & $0.1$ & $0.0077$ & $9.2$ & $1.8$ & \textquotedbl \\
3 & Ms7.0Ma1.8  & $0.1$ & $0.01$     & $7.0$ & $1.8$ & \textquotedbl \\
4 & Ms4.3Ma1.5  & $0.1$ & $0.025$   & $4.3$ & $1.5$ & \textquotedbl \\
5 & Ms3.1Ma1.7  & $0.1$ & $0.05$     & $3.1$ & $1.7$ & \textquotedbl \\
6 & Ms2.4Ma1.9  & $0.1$ & $0.1$    & $2.4$ & $1.9$ & \textquotedbl \\
7 & Ms0.8Ma1.7  & $0.1$ & $0.7$    & $0.8$ & $1.7$ & Transonic and super-\alf ic \\
8 & Ms0.5Ma1.7  & $0.1$ & $2$       & $0.5$ & $1.7$ & Subsonic and super-\alf ic \\
\hline
9   & Ms9.9Ma0.5    & $1$ & $0.0049$ & $9.9$ & $0.5$ & Supersonic and sub-\alf ic \\
10 & Ms7.9Ma0.5    & $1$ & $0.0077$ & $7.9$ & $0.5$ & \textquotedbl \\
11 & Ms6.8Ma0.5  & $1$ & $0.01$   & $6.8$ & $0.5$ & \textquotedbl \\
12 & Ms4.5Ma0.6  & $1$ & $0.025$ & $4.5$ & $0.6$ & \textquotedbl \\
13 & Ms3.2Ma0.6  & $1$ & $0.05$   & $3.2$ & $0.6$ & \textquotedbl \\
14 & Ms2.4Ma0.7  & $1$ & $0.1$     & $2.4$ & $0.7$ & \textquotedbl \\
15 & Ms0.9Ma0.7    & $1$ & $0.7$     & $0.9$ & $0.7$ & Transonic and sub-\alf ic \\
16 & Ms0.5Ma0.7  & $1$ & $2$       & $0.5$ & $0.7$ & Subsonic and sub-\alf ic \\
\hline
\end{tabular}
\end{table*}

\section{Production of Synthetic Polarization Maps}
\label{synthetic}

\begin{figure*}
\begin{center}
\includegraphics[scale=0.24]{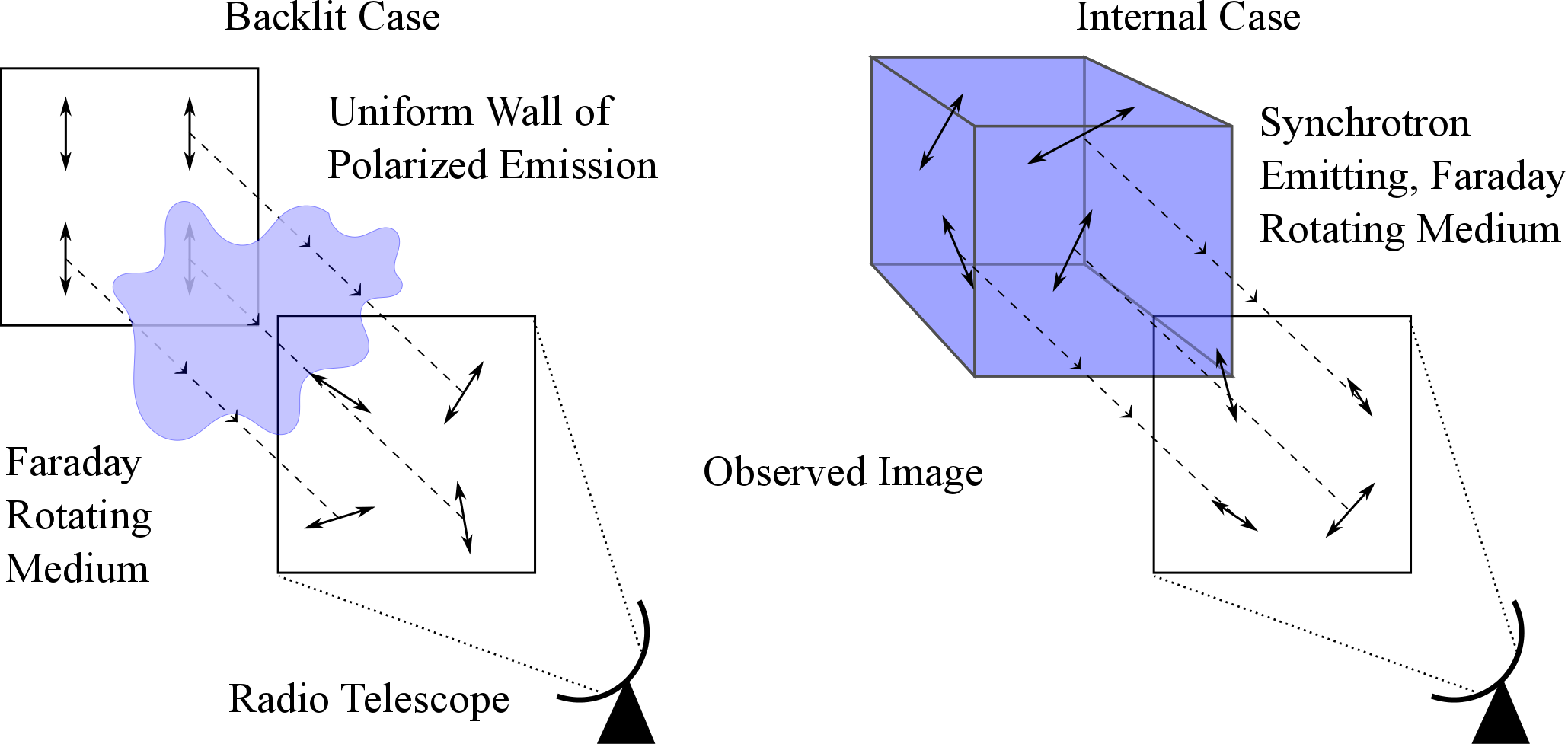}
\caption{A diagram illustrating the differences between the cases of backlit and internal emission propagating through a Faraday rotating medium. The backlit case is shown on the left, where a wall of uniform polarized emission propagates through a Faraday rotating medium, causing the observed polarization angles to vary across the image, although the polarization intensity remains uniform. The internal case is shown on the right, where polarized emission is radiated from each point within the emitting, Faraday rotating medium. In this case, the initial polarization angle and polarization intensity is determined by the magnetic field at the point of emission, and this emission is rotated as it propagates through the medium. Polarization from different depths within the cube destructively interferes, causing the observed polarization intensity and polarization angle to vary across the image.}
\label{back_vs_int}
\end{center}
\end{figure*}

We define two methods used to derive synthetic maps of Stokes $Q$ and $U$ for our simulations, which are illustrated in Fig. \ref{back_vs_int}. In the `backlit' case, the turbulent cube is illuminated from behind by a uniform wall of polarized synchrotron emission. We assume that this wall has unit polarization intensity and uniform polarization angle equal to zero everywhere across it. This corresponds to $Q=-1$, and $U=0$ everywhere. As the emission passes through the turbulent cube, the emission is Faraday rotated according to Eq. \ref{Faraday_rotation}. The final observed polarisation angle is given by $\psi(x,y,\lambda^2) = \text{RM}(x,y) \, \lambda^2$, and so the observed Stokes $Q$ and $U$ are given by $Q(x,y,\lambda^2) = P \cos 2 \psi(x,y,\lambda^2)$ and $U(x,y,\lambda^2) = P \sin 2 \psi(x,y,\lambda^2)$, where we include the polarization intensity to show that it is uniform across the image, and independent of wavelength. The backlit case represents the simplest way in which polarized emission can propagate through a turbulent medium, against which we can compare the results obtained for the more realistic scenario of emission originating within the turbulent volume, which we refer to as the `internal' case.

In the internal case, the polarized synchrotron emission arises from within the cube, and the emissivity at a pixel is given by the integrand of Eq. \ref{sync_inten}. The polarization emissivity is found by multiplying this by the fractional polarization (Eq. \ref{frac_pol}), and the intrinsic polarization angle at this pixel is determined by calculating the direction of the magnetic field perpendicular to the line of sight (in the plane of the sky), and adding $90$ degrees. Starting from the front of the cube, we calculate the intrinsic polarization intensity and polarization angle at each pixel, and perform Faraday rotation due to the material in front of the current slice. We then calculate the Stokes $Q$ and $U$ that would be observed from this slice, based on the intrinsic polarization intensity and the rotated polarization angle. We then move to the next slice, and increment the rotation measure by the product of the electron density and the magnetic field parallel to the line of sight. We again calculate the Stokes $Q$ and $U$ that would be observed from this slice after Faraday rotation by material in front of the slice, and add these values to the total Stokes $Q$ and $U$. This process is then repeated for all slices along the line of sight, until the polarized emission from each slice has been added together. This process naturally accounts for the wavelength independent depolarization that arises along the line of sight, due to emission with different intrinsic polarization angles superimposing, and for the depolarization due to differential Faraday rotation, namely that emission from different depths within the cube is rotated by different amounts, leading to interference of the polarization vectors.

To ensure that diagnostics calculated for the cases of backlit and internal emission can be directly compared, we normalize the polarization intensity for the case of internal emission. We perform this normalization by dividing the observed, total Stokes $Q$ and $U$ by the average polarization intensity that would be observed if there was no depolarization, called $P^*$. We calculate $P^*$ by integrating the polarization emissivity at each pixel along the line of sight, and then averaging this over the image to obtain a constant. The normalized complex polarization vector that we calculate, $\boldsymbol{P}_n$, is then given by
\begin{equation}
\boldsymbol{P}_n = \frac{ \int_{0}^{L} B_{\perp}^{1-\alpha} \exp [2i (\psi_0 + \text{FD} \, \lambda^2)] \mathrm{d} z }{ \left< \int_{0}^{L} B_{\perp}^{1-\alpha} \mathrm{d} z \right> }, \label{norm_pol_spec}
\end{equation}
where $z=L$ corresponds to the slice at the back of the simulation cube, and we omit dependence on the Cartesian coordinate system. By performing this normalization, we ensure that the total amount of energy injected into polarized emission is the same for the backlit and internal emission cases, and that polarization diagnostics calculated for the cases of backlit and internal emission can be directly compared.

What this formula demonstrates is that the normalized complex polarization depends upon the spectral index $\alpha$. This is a form of spectral depolarization that arises because a more negative spectral index causes the contrast between regions of high and low magnetic field to be enhanced, such that the observed polarization is mostly determined by the regions of strongest magnetic field. This affects the interference of polarization vectors along each sightline, and hence the observed polarization intensity.

\begin{figure*}
\begin{center}
\includegraphics[scale=0.78]{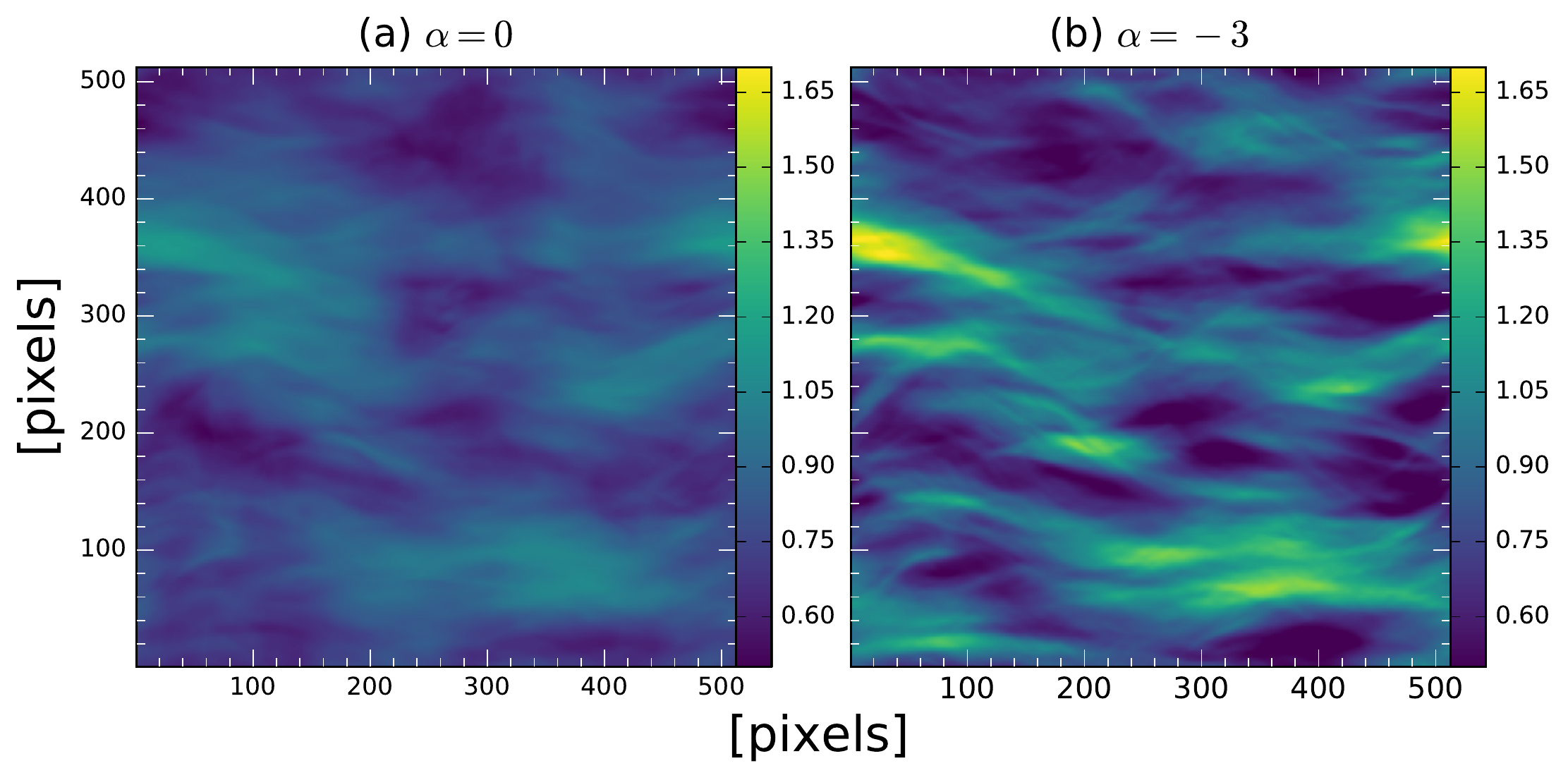}
\caption{The normalized polarization intensity (dimensionless) for the Ms0.5Ma0.7 simulation, and for internal emission, for two different spectral indices. a) The polarization intensity for a spectral index $\alpha = 0$. b) The polarization intensity for a spectral index $\alpha = -3$. For both images, a line of sight along the $y$ axis is used, and the observing frequency is $1.4$ GHz.}
\label{spec_ind_diff}
\end{center}
\end{figure*}

We studied the influence of the spectral index on our synthetic observations by calculating the polarization intensity and polarization angle for the Ms0.5Ma0.7 simulation for different spectral indices, for the case of internal emission. Example images are shown in Fig. \ref{spec_ind_diff}, for spectral indices of $0$ (left) and $-3$ (right). We find that there are significant changes in the polarization intensity between spectral indices of $0$ and $-3$, but there was little change in the polarization angle. However, for spectral index values between $-0.5$ and $-1.5$, typical of the Milky Way (see \citealt{Herron2016}, and references therein), we found that the polarization intensity changes by at most $15\%$, and in general the polarization intensity and polariation angle do not change very much. Following \cite{Herron2016}, we choose a spectral index of $-1$ for all of our synthetic observations, as this value is similar to that observed in the Galaxy.

A consequence of our chosen normalization (Eq. \ref{norm_pol_spec}) is that we have removed the wavelength dependence of the synchrotron emissivity, so that the wavelength dependence of the complex polarization is only caused by differential Faraday rotation, and this will affect the derivatives with respect to wavelength that we will calculate. As this normalization cannot be applied to observed polarization maps, it is not possible to directly compare the derivatives with respect to wavelength that we calculate for our normalized polarization maps to observed polarization maps. To be able to compare the wavelength derivatives that we calculate for our simulated polarization maps to observations, it is necessary to be able to scale the wavelength derivative of the normalized polarization map, to the wavelength derivative of the original (un-normalized) polarization map, which can be directly compared to observations. To check that it will be possible to convert derivatives with respect to wavelength calculated for the original complex polarization and the normalized complex polarization, we derived the following formula linking the two:
\begin{equation}
\frac{\mathrm{d} \boldsymbol{P}}{\mathrm{d} \lambda^2}(x,y,\lambda^2) = - \frac{\alpha}{2 \lambda^2} \boldsymbol{P}(x,y,\lambda^2) + P^* \frac{\mathrm{d} \boldsymbol{P}_n}{\mathrm{d} \lambda^2}(x,y,\lambda^2). \label{pol_norm}
\end{equation}
Eq. \ref{pol_norm} shows that the wavelength derivative of the original complex polarization (on the left hand side), can be calculated from the dependence of the complex polarization on wavelength due to the synchrotron emissivity (first term on the right), and the wavelength dependence of the normalized complex polarization $\boldsymbol{P}_n$ (second term on the right), multiplied by the polarization intensity that would be observed in the absence of depolarization, $P^*$. Using this equation, it is possible to convert between derivatives with respect to wavelength that were calculated for the original polarization or the normalized polarization for our simulations, and so it is valid to just examine the normalized polarization, which provides a more convenient means of studying the influence of differential Faraday rotation.

Synthetic observations of Stokes $Q$ and $U$ were calculated for all of our simulations, for lines of sight along each axis of the simulation. For the case of backlit emission, we assumed a frequency of $1.4$ GHz. Only one frequency is required, since what is observed at other frequencies can be easily calculated from the rotation measure, as we do not include the effects of beam depolarisation. For the case of internal emission, we chose $50$ frequencies between $0.5$ GHz and $2$ GHz, equally separated in wavelength squared space. To calculate spatial derivatives of $Q$ and $U$ at a pixel, the gradient is calculated between the adjacent pixels. At the boundary of the image, the spatial derivatives are calculated from the gradient between the pixel itself and the adjacent pixel. For wavelength derivatives, a similar method is applied to the adjacent wavelength slices of the data cube.

\begin{figure*}
\begin{center}
\includegraphics[scale=0.78]{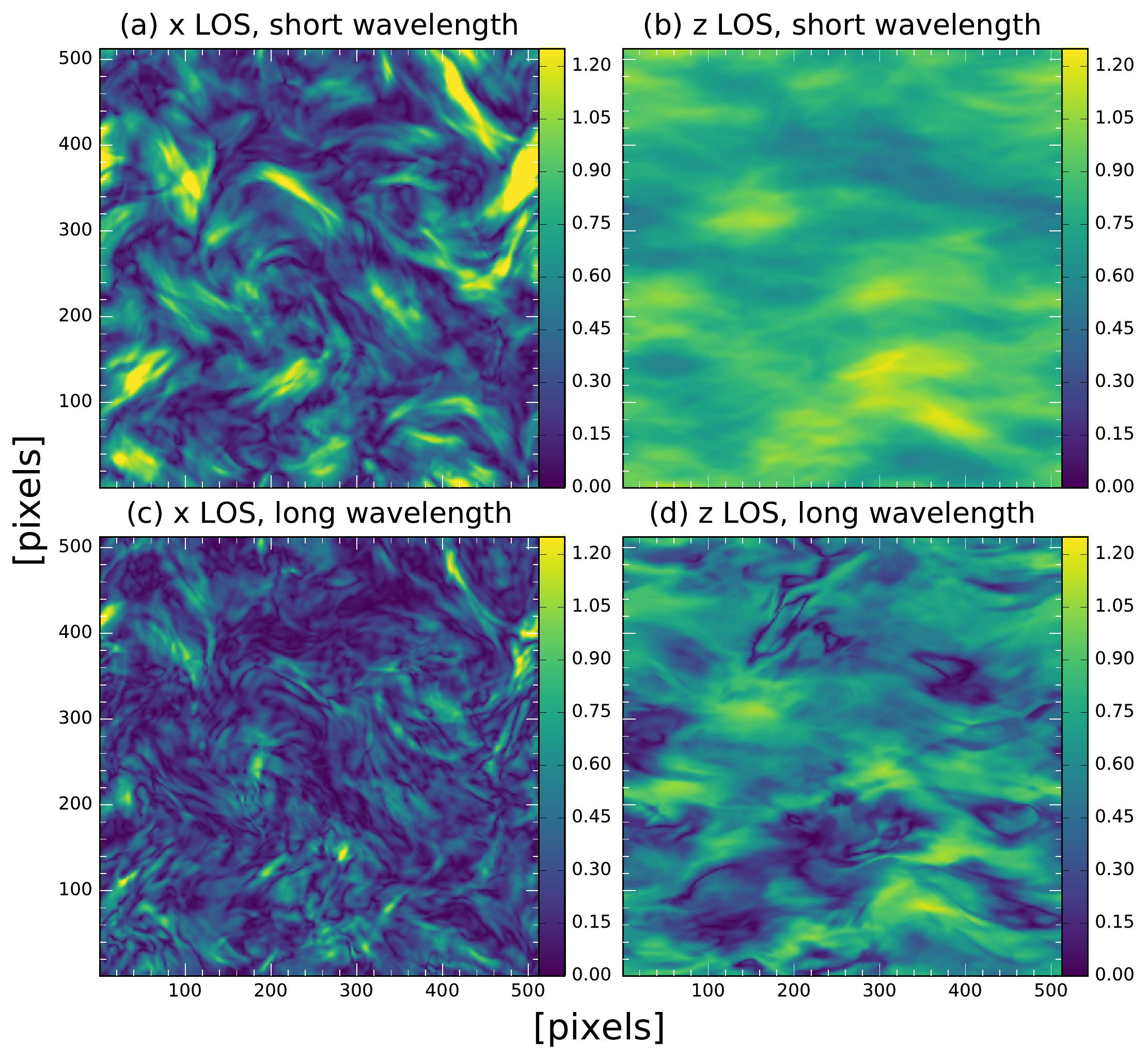}
\caption{The normalized polarization intensity (dimensionless) observed for the Ms0.9Ma0.7 simulation, for internal emission, and different lines of sight and observing wavelengths. A line of sight along the $x$ axis (parallel to the mean magnetic field) is used on the left, and a line of sight along the $z$ axis (perpendicular to the mean magnetic field) is used on the right. The observing frequency is $2$ GHz on the top row, and $0.5$ GHz on the bottom row.}
\label{pol_inten_examples}
\end{center}
\end{figure*}

In Fig. \ref{pol_inten_examples} we show example polarization intensity images for the Ms0.9Ma0.7 simulation, for internal emission, lines of sight parallel (left) and perpendicular (right) to the mean magnetic field, at short (top) and long (bottom) wavelengths. We find that there is more small-scale structure at long wavelengths, due to the greater degree of Faraday rotation and depolarization, and also find that structures tend to be parallel to the mean magnetic field if our line of sight is perpendicular to the field.

\section{Polarization Gradient for Internal Emission}
\label{internal_grad}

Previously, \cite{Burkhart2012} studied the polarization gradient for the case of backlit emission. In this section, we explore the properties of the polarization gradient and generalized polarization gradient for the cases of backlit and internal emission, and lines of sight parallel and perpendicular to the mean magnetic field.

For backlit emission, we find that the polarization gradient traces spatial variations in the magnetoionic medium for all lines of sight. We also find that for lines of sight perpendicular to the mean magnetic field, the polarization gradient structures tend to align with the magnetic field, provided that the simulation is subsonic and sub-Alfv\'enic. As the sonic Mach number of the simulation increases, there is an increase in the amount of small scale structure, and a clumpy topology may be seen.

For internal emission, we similarly find that polarization gradient structures are aligned with the mean magnetic field for subsonic simulations with a strong magnetic field perpendicular to the line of sight. However, the polarization gradient is only sensitive to spatial variations in the magnetoionic medium across the image for subsonic simulations. For supersonic simulations, depolarization due to differential Faraday rotation becomes important, and in this case the polarization gradient is dominated by variations in the degree of differential Faraday rotation.

\begin{figure*}
\begin{center}
\includegraphics[scale=0.78]{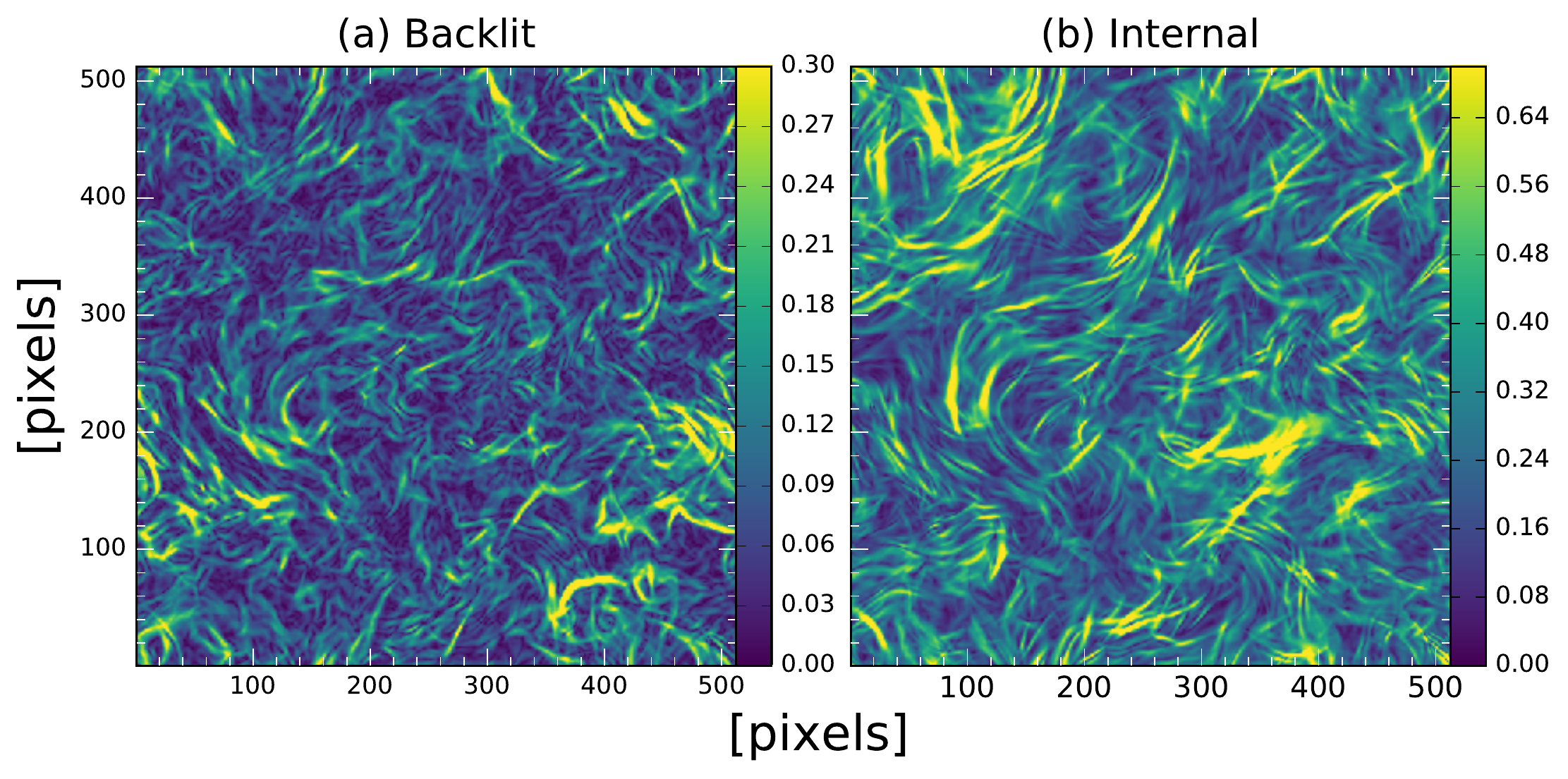}
\caption{Polarization gradient images observed for the cases of backlit emission (a, left), and internal emission (b, right), for the Ms0.5Ma0.7 simulation, a line of sight along the $x$ axis, and a frequency of $1.4$ GHz. Units are pc$^{-1}$. Different color scalings are used for the images.}
\label{pol_grad_back_int}
\end{center}
\end{figure*}

In Fig. \ref{pol_grad_back_int} we show the polarization gradient images for the Ms0.5Ma0.7 simulation, a line of sight parallel to the mean magnetic field, for the case of backlit (left) and internal (right) emission, both at a frequency of $1.4$ GHz. We find that the images produced for the backlit and internal cases display structures of very different morphology, with the internal case exhibiting filaments that are straighter than those seen in the backlit case. Hence, any attempt to constrain the physical properties of an observed turbulent region by using observed statistics of polarization diagnostics must consider whether the turbulent volume is backlit by polarized emission, or polarized emission is generated within the volume. 

\begin{figure*}
\begin{center}
\includegraphics[scale=0.8]{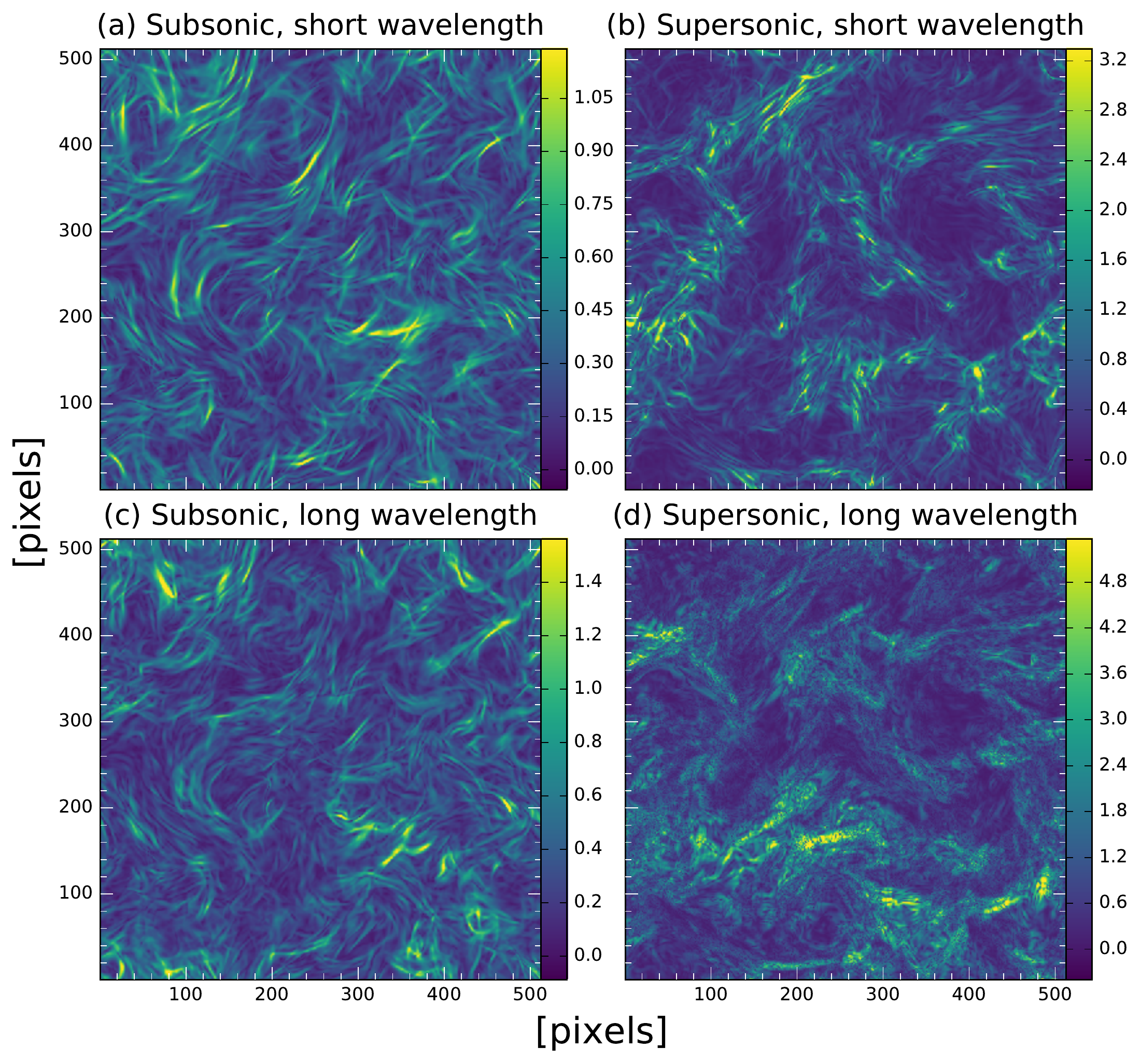}
\caption{Polarization gradient images for a subsonic simulation (left column, Ms0.5Ma0.7) and supersonic simulation (right column, Ms3.2Ma0.6), for internal emission, and a line of sight along the $x$ axis. The images in the top row were produced with a frequency of $2$ GHz, and the images in the bottom row were produced with a frequency of $0.5$ GHz. Units are pc$^{-1}$. Different color scalings are used for the images.}
\label{pol_grad_sub_super}
\end{center}
\end{figure*}

In Fig. \ref{pol_grad_sub_super} we show the polarization gradient for the Ms0.5Ma0.7 simulation (left) and the Ms3.2Ma0.6 simulation (right), for the case of internal emission, a line of sight parallel to the mean magnetic field, at two different wavelengths, to demonstrate the sensitivity of polarization gradient structures to the sonic Mach number. We observe that supersonic simulations have much smaller scale structure than subsonic simulations. This gives supersonic simulations a clumpier appearance, with larger contrast between regions of large and small polarization gradient, compared to subsonic simulations. As the observing wavelength increases, all simulations exhibit more small-scale structure, although this is more significant for simulations that are supersonic, or have a strong magnetic field parallel to the line of sight. In particular, for supersonic simulations the small-scale polarization gradient structure begins to appear as though it is superimposed on large-scale regions of large polarization gradient. Polarization gradient structures are hence sensitive to the sonic Mach number for the case of internal emission, and the genus, which was shown by \cite{Burkhart2012} to be sensitive to the sonic Mach number for the case of backlit emission, may also be useful for the case of internal emission, provided the observing wavelength is taken into account.

\begin{figure*}
\begin{center}
\includegraphics[scale=0.7]{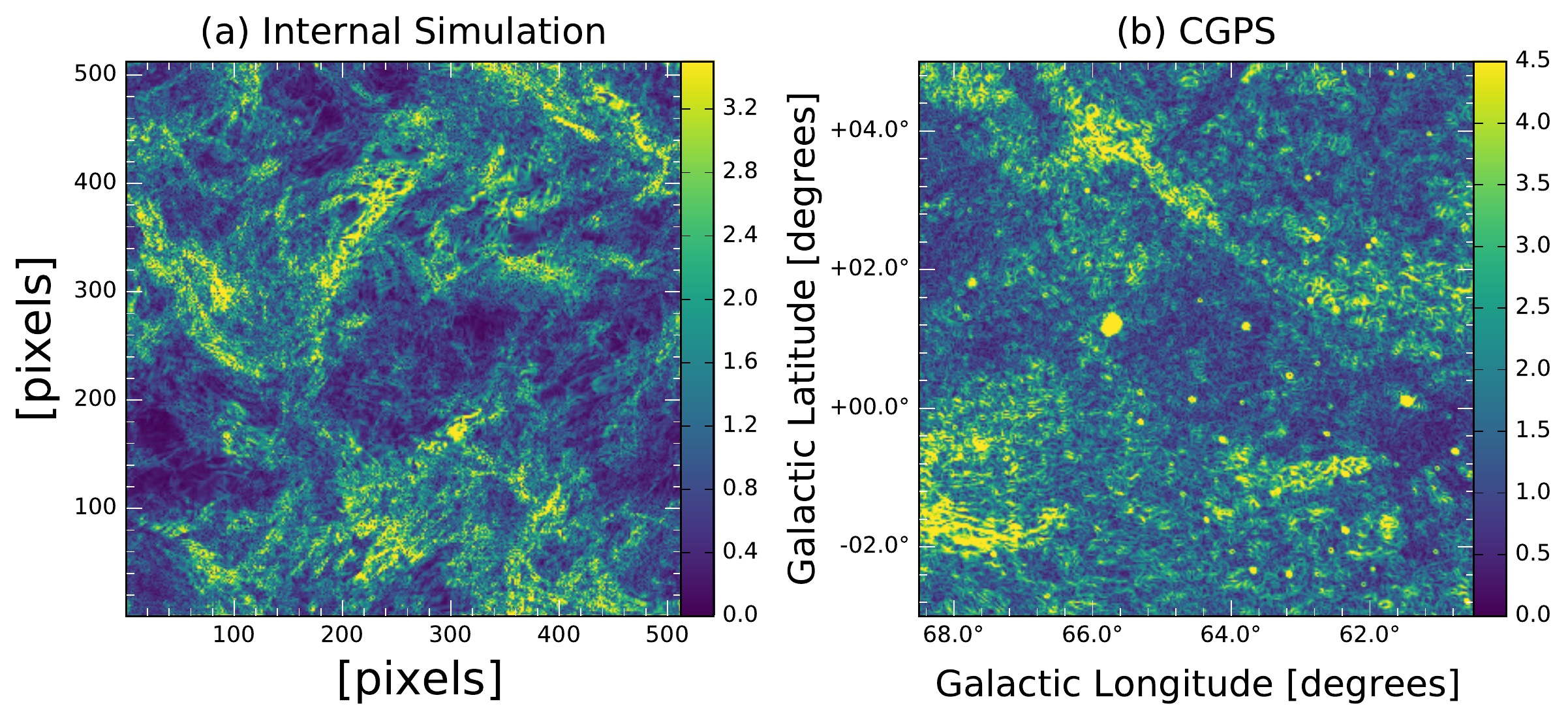}
\caption{Polarization gradient images for the Ms7.0Ma1.8 simulation (a, left) and a portion of the Canadian Galactic Plane Survey (CGPS) toward a Galactic longitude of $65^{\circ}$ (b, right, see \citealt{Herron2017b} for more information). The simulated image was produced with internal emission, a line of sight along the $x$ axis, and a frequency of $0.5$ GHz, and values are given in units of pc$^{-1}$. The CGPS image has an angular resolution of $150$ arc seconds, and values are in units of K per degree. Different color scalings are used for the images.}
\label{pol_grad_cgps}
\end{center}
\end{figure*}

We notice that the appearance of the polarization gradient at long wavelengths for supersonic simulations, namely small-scale structure superimposed on large-scale features, is reminiscent of the polarization gradient features seen in the Canadian Galactic Plane Survey (CGPS, \citealt{Landecker2010}) at low longitudes (see \citealt{Herron2017b} for the full polarization gradient images). In Fig. \ref{pol_grad_cgps} we compare the polarization gradients synthesized for the Ms7.0Ma1.8 simulation, for a line of sight parallel to the mean magnetic field, internal emission, and a frequency of $0.5$ GHz, to the polarization gradient observed in the CGPS toward a Galactic longitude of $65^{\circ}$. We find that the CGPS gradient image also shows small-scale structure that appears to be superimposed on large-scale structure, which may indicate that the turbulence observed in this region of the CGPS is supersonic, and that the observed radiation is predominantly emitted within a Faraday rotating medium. We do not believe that these structures are noise, as the maps of $Q$ and $U$ were smoothed to obtain good signal-to-noise prior to producing this image.

\begin{figure*}
\begin{center}
\includegraphics[scale=0.8]{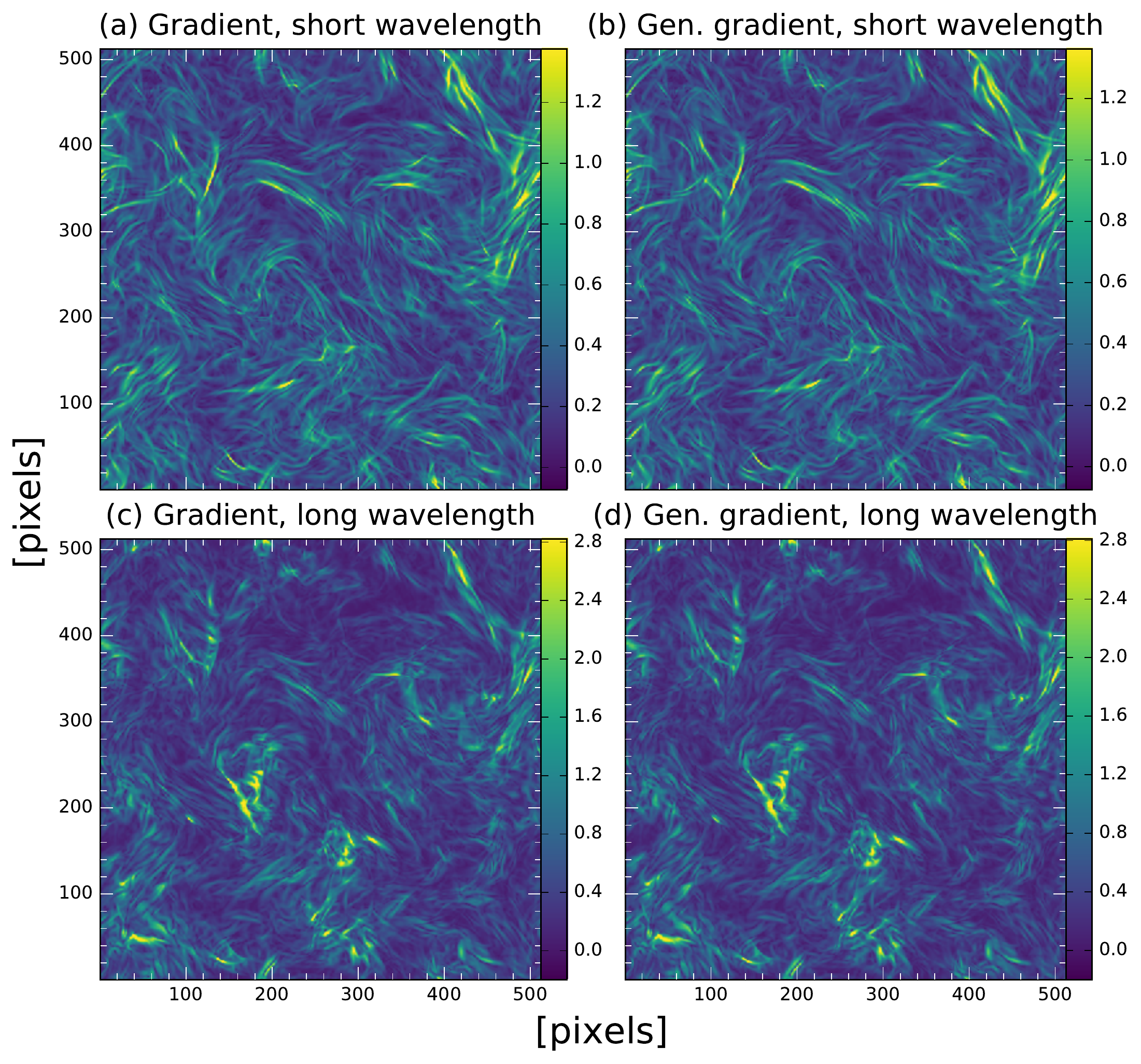}
\caption{The polarization gradient (left column) and generalized polarization gradient (right column) for the Ms0.9Ma0.7 simulation, internal emission, and a line of sight along the $x$ axis. A frequency of $2$ GHz was used for the images in the top row, and $0.5$ GHz for the images in the bottom row. Units are pc$^{-1}$. Different color scalings are used for the images.}
\label{pol_grad_vs_general}
\end{center}
\end{figure*}

Finally, in Fig. \ref{pol_grad_vs_general} we compare the polarization gradient (left) to the generalized polarization gradient (right) for internal emission, to examine similarities in their structures, as we have shown in Paper I that they are identical for backlit emission. This comparison is performed for the Ms0.9Ma0.7 simulation, at two different wavelengths. We find that there is very little difference between the polarization gradient and the generalized polarization gradient, for any wavelength. This is also true for any simulation, and any line of sight, and so the generalized polarization gradient should also be sensitive to spatial variations in the magnetoionic medium, and the structure seen in images of the generalized polarization gradient should be sensitive to the sonic Mach number of the turbulent region observed.

\section{Radial and Tangential Components of the Directional Derivative}
\label{rad_tang_comp}

By calculating the radial and tangential components of the directional derivative, it is possible to quantify how changes in polarization intensity and polarization angle contribute to the generalized polarization gradient. The maximum amplitude of the radial component measures the maximal contribution of changes in polarization intensity to directional derivative, and likewise the maximum amplitude of the tangential component measures the maximal contribution of changes in the polarization angle. Together, these diagnostics may allow us to study individual features seen in maps of the generalized polarization gradient. In this Section we discuss how the radial and tangential components compare to the generalized polarization gradient, and possible uses of these diagnostics.

\begin{figure*}
\begin{center}
\includegraphics[scale=0.7]{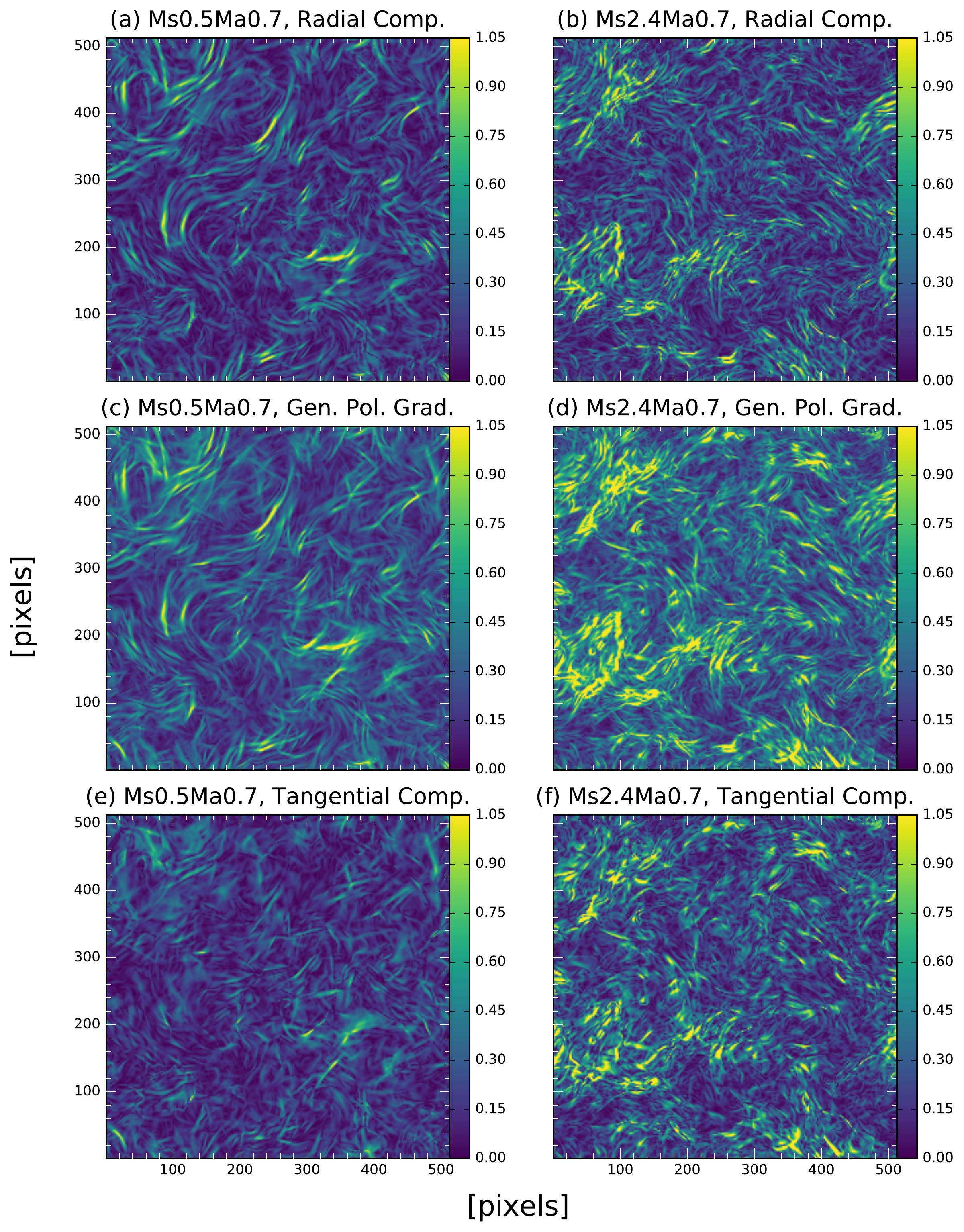}
\caption{A comparison of the maximum amplitude of the radial (top row) and tangential (bottom row) components of the directional derivative to the generalized polarization gradient (middle row), for the subsonic Ms0.5Ma0.7 simulation (left column) and supersonic Ms2.4Ma0.7 simulation (right column). These images were produced for internal emission, a line of sight along the $x$ axis, at a frequency of $2$ GHz. The units of all figures are pc$^{-1}$.}
\label{rad_tang_direc_comparison}
\end{center}
\end{figure*}

In Fig. \ref{rad_tang_direc_comparison} we show the maximum amplitudes of the radial (top) and tangential (bottom) components of the directional derivative, and the generalized polarization gradient (middle) for the Ms0.5Ma0.7 (left) and Ms2.4Ma0.7 (right) simulations, a line of sight along the $x$ axis, internal emission, at a frequency of $2$ GHz. For the Ms0.5Ma0.7 simulation, it is clear that the generalized polarization gradient is dominated by the radial component, and changes in polarization intensity, as these two images display similar structures. However, there are some features that are solely caused by changes in polarization angle, for example the bright filament in the bottom left of the image of the tangential component appears in the image for the generalized polarization gradient, but only has a faint counterpart in the image for the radial component.

For the Ms2.4Ma0.7 simulation, we find that the generalized polarization gradient is most similar to the tangential component, based on the brightness of these two quantities, although the features seen in the radial component are also very similar to the generalized polarization gradient. There are some features of the generalized polarization gradient that are predominantly caused by changes in polarization intensity, for example the two bright filaments toward the bottom of the radial component image, whereas other features are predominantly caused by the changes in polarization angle, such as the bright filament in the top right of the image.

\begin{figure*}
\begin{center}
\includegraphics[scale=0.78]{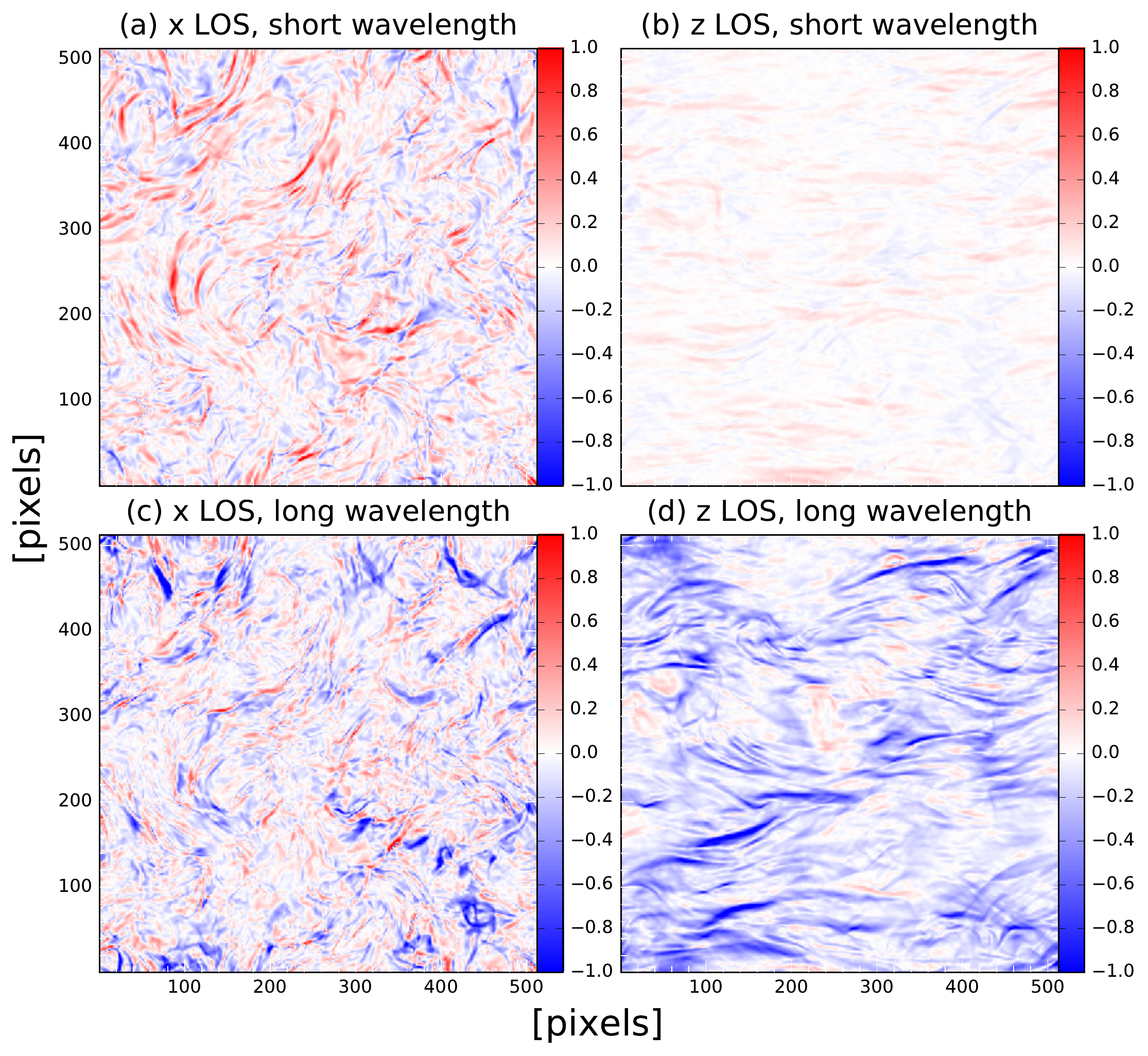}
\caption{The difference between the maximum amplitudes of the radial and tangential components of the directional derivative, observed for the Ms0.5Ma0.7 simulation, for internal emission, and different lines of sight and observing wavelengths. A line of sight along the $x$ axis (parallel to the mean magnetic field) is used on the left, and a line of sight along the $z$ axis (perpendicular to the mean magnetic field) is used on the right. The observing frequency is $2$ GHz on the top row, and $0.5$ GHz on the bottom row. Red denotes regions where the radial component, and changes in polarization intensity, dominate, and blue denotes regions where the tangential component, and changes in polarization angle dominate.}
\label{rad_tang_diff_sims}
\end{center}
\end{figure*}

A convenient way of examining whether the radial or tangential component dominates the generalized polarization gradient is to calculate the difference between these components. In Fig. \ref{rad_tang_diff_sims} we show the result of subtracting the maximum value of the tangential component of the directional derivative from the maximum amplitude of the radial component, for the Ms0.5Ma0.7 simulation, internal emission, lines of sight along the $x$ (left) and $z$ (right) axes, at frequencies of $2$ GHz (top) and $0.5$ GHz (bottom). Red corresponds to areas dominated by the radial component, and blue corresponds to areas dominated by the tangential component.

We find that if the component of the magnetic field in the plane of the sky is small, as is the case for a line of sight parallel to the mean magnetic field, then red and blue regions are intermixed. If the component of the magnetic field perpendicular to the line of sight is large, then the image tends to be dominated by either red (if there is little Faraday rotation) or blue (if there is significant Faraday rotation), with few features of the other colour. In general, the tangential component becomes larger as the wavelength increases, causing these images to have strong blue features. This is likely because the amount of Faraday rotation is larger at longer wavelengths, so that there are larger differences in the observed polarization angle.

\begin{figure*}
\begin{center}
\includegraphics[scale=0.85]{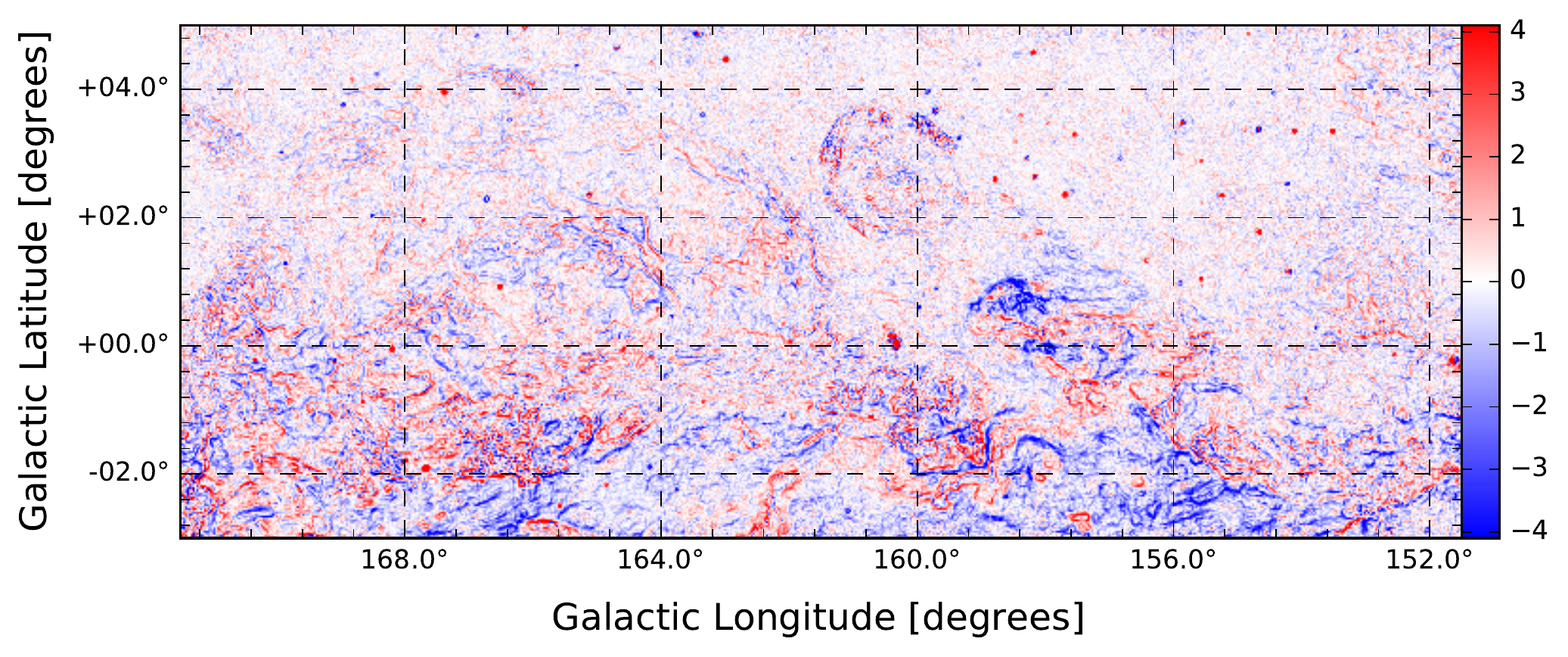}
\caption{The difference between the maximum amplitudes of the radial and tangential components of the directional derivative for a portion of the Canadian Galactic Plane Survey at an angular resolution of $150$ arc seconds, in units of K per degree (see \citealt{Herron2017b} for more information). Red denotes regions where the radial component, and changes in polarization intensity, dominate, and blue denotes regions where the tangential component, and changes in polarization angle dominate.}
\label{rad_tang_diff_cgps}
\end{center}
\end{figure*}

In Fig. \ref{rad_tang_diff_cgps}, we show the difference between the maximum amplitudes of the radial and tangential components of the directional derivative for a section of the polarization gradient image of the Canadian Galactic Plane Survey, produced by \cite{Herron2017b}. We find that, in general, red and blue regions appear to be intermixed, such that regions dominated by changes in polarization intensity and polarization angle alternate across the image. The most prominent exception to this is shown in Fig. \ref{rad_tang_diff_cgps}, where there is an extended area between Galactic longitudes of $152^{\circ} < \ell < 168^{\circ}$ and Galactic latitudes of $-3^{\circ} < b < -1^{\circ}$, for which changes in the polarization angle dominate.

This extended region of strong tangential component may imply that there is a strong, large-scale magnetic field perpendicular to the line of sight in this area, or that there is a large-scale gradient in the rotation measure across this area, that causes the observed polarization angle to have strong spatial dependence. Conversely, areas with intermixed blue and red filaments may imply that small-scale turbulence is responsible for the observed polarimetric features in this area, without a strong component of the magnetic field in the plane of the sky. This is because small-scale turbulence can cause spatial changes in polarization intensity or polarization angle due to differential Faraday rotation.

The radial and tangential components of the directional derivative can hence provide qualitative insight on whether observed polarimetric features are produced by turbulence, or large-scale Galactic features, in addition to conveying whether polarization gradient structures are caused by changes in polarization intensity or polarization angle.

\section{Methods to Distinguish Backlit and Internal Emission}
\label{back_int}
As shown in Section \ref{internal_grad}, whether the observed polarized emission comes from within or behind a turbulent magnetoionic region has a strong influence on how we interpret polarimetric data, and also on the properties of the turbulent region that we might try to infer from statistics of polarimetric diagnostics. Recently, \cite{Sun2014} introduced a method of distinguishing between backlit and internal emission that involves calculating the structure function of the polarization intensity, and the complex polarization, and comparing the slopes of these structure functions. They found that if the structure function of the complex polarization has a flatter slope than the structure function of polarization intensity, then the emission is caused by foreground Faraday screens, corresponding to our backlit case. If the slopes are similar, then the emission is intrinsic to the turbulent medium, corresponding to our internal case. In this Section, we derive complementary methods of distinguishing between backlit and internal emission.

One method of distinguishing between backlit and internal emission involves the radial component of the directional derivative, and the radial component of the polarization wavelength derivative. For uniform backlit emission, the polarization intensity should be uniform across the image, and independent of wavelength (ignoring the dependence of synchrotron intensity on wavelength due to its spectral index). This means that the radial component of the directional derivative should be identically equal to zero, as should the radial component of the wavelength derivative, i.e.
\begin{equation}
\frac{\partial \boldsymbol{P}}{\partial s}_{\text{rad}} = 0 \quad \text{and} \quad \frac{\partial \boldsymbol{P}}{\partial \lambda^2}_{\text{rad}} = 0 \label{rad_comp_zero}
\end{equation}
respectively, where $s$ denotes distance in the image plane. If either of these radial components is non-zero, then this may imply that the emission is generated within the turbulent region, or that beam depolarization, where polarization vectors within the telescope beam destructively interfere, is important.

Another method involves the gradients of Stokes $Q$ and $U$. For backlit emission, the gradients of $Q$ and $U$ should be in the same direction, namely in the direction of the gradient of the polarization angle. This means that the cross product between the gradients of Stokes $Q$ and $U$ should be identically zero for uniform, backlit emission. This method is equivalent to measuring the difference between the polarization gradient and the generalized polarization gradient, as the generalized polarization gradient only differs from the polarization gradient due to a term that is equal to the amplitude of the cross product of the gradients of Stokes $Q$ and $U$ (see Eqs. 2 and 15 of Paper I).

The polarization directional curvature and wavelength curvature provide alternative methods for distinguishing between backlit and internal emission. In the following, we assume that the interferometric data is complemented by single dish data, so that the true polarisation intensity is measured. For the case of uniform, backlit emission, observable polarization values lie on a circle of radius equal to the polarization intensity, centred on the origin of the \QU plane. This means that as we move across the image, the observed polarization vector traces out a circular arc of radius $P$, and whose curvature must be $1/P$. Similarly, if we examine how the polarization vector changes with wavelength at a pixel, a circular arc of radius $P$ is traced.

It follows that for backlit emission, the polarization directional curvature and wavelength curvature should be equal to $1/P$ at every pixel of the image, at every wavelength, provided that it is valid to calculate the curvature at that pixel. It is valid to calculate the directional curvature if the directional derivative is non-zero in the specified direction, and it is valid to calculate the wavelength curvature if the wavelength derivative is non-zero at the specified pixel. As mentioned in Paper I, calculating the curvature in the direction that maximizes the directional derivative ensures that the directional curvature is calculated at every pixel where it is valid to do so, and hence this diagnostic provides a convenient method of examining whether we have observed backlit or internal emission. We caution, however, that we have not yet considered how beam depolarization will affect the directional curvature or the wavelength curvature, and that the wavelength dependence due to the synchrotron spectral index must be taken into account before using the wavelength curvature.

\begin{figure*}
\begin{center}
\includegraphics[scale=0.8]{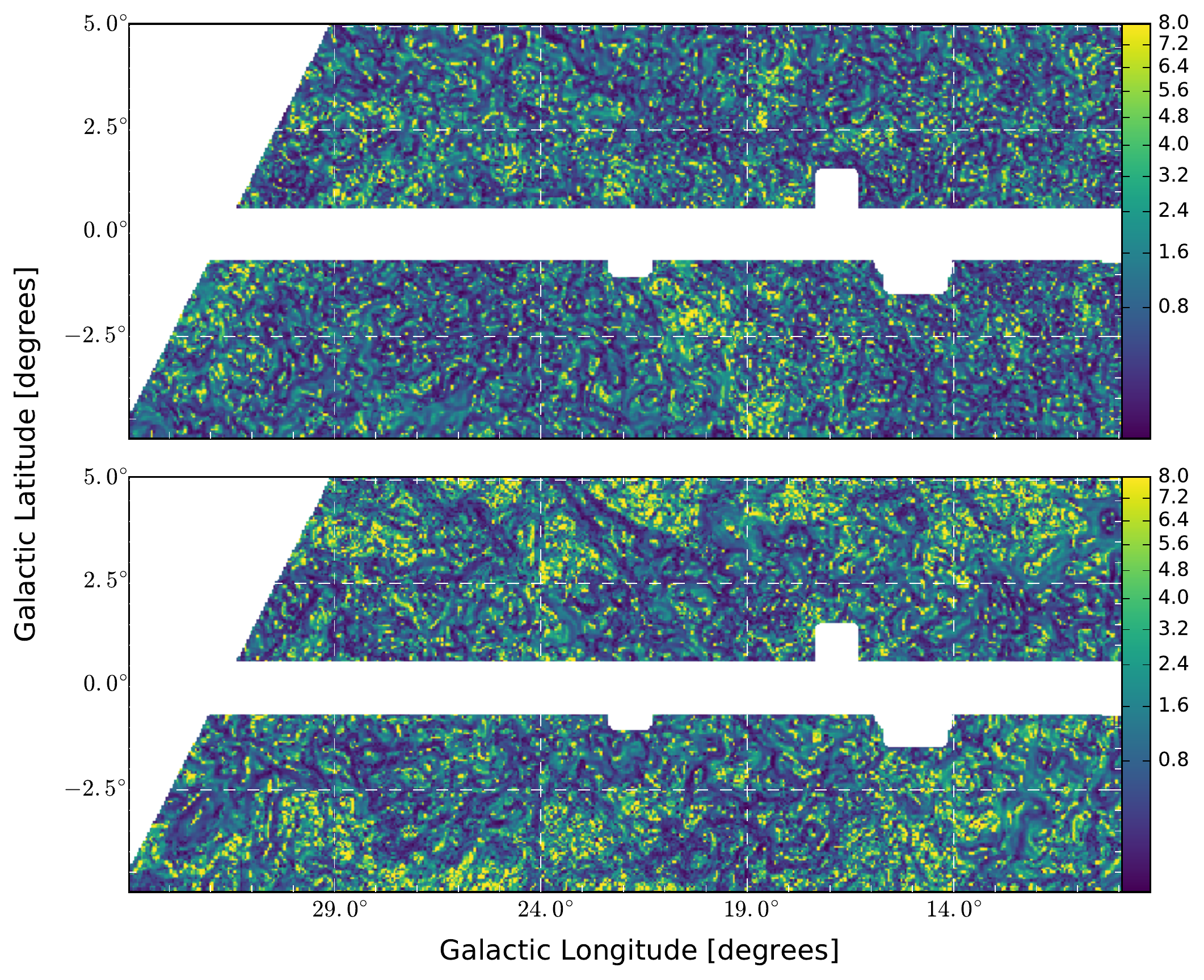}
\caption{The curvature in the direction that maximizes the directional derivative multiplied by the polarization intensity, for the $2.3$ GHz S-band Polarization All Sky Survey (top) and $4.8$ GHz Urumqi telescope (bottom) observations used by \cite{Sun2014}. The units for the curvature are mK$^2$ per square degree. Regions of large curvature differ for the two frequencies.}
\label{Xiaohui_curv}
\end{center}
\end{figure*}

In Fig. \ref{Xiaohui_curv} we calculate the polarization curvature in the direction that maximizes the directional derivative, for the $2.3$ GHz (top, S-band Polarization All Sky Survey, \citealt{Carretti2010, Carretti2013}) and $4.8$ GHz (bottom, Sino-German $\lambda 6$ cm survey, \citealt{Sun2011}) data used by \cite{Sun2014}. In these images we have multiplied the polarization curvature by the polarization intensity, so that we expect to see a value of $1$ across the image, if the observations are of backlit emission. 

\cite{Sun2014} found that the polarized emission they observed at $4.8$ GHz was internal, and that the polarized emission observed at $2.3$ GHz was backlit. For both images, we find that the directional curvature multiplied by polarization intensity is not equal to $1$ over the image, in general. This would suggest that the observed polarized emission is generated within the turbulent volume, however our method does not take into account beam depolarization, and so should be treated with caution. Although not shown here, we also note that the directional curvature features observed at $2.3$ GHz tend to be brighter toward the Galactic plane, whereas the features observed at $4.8$ GHz tend to be brighter away from the Galactic plane. A discussion of this is beyond the scope of this paper. 

Finally, we emphasise that the methods we have developed to distinguish between the cases of backlit and internal emission work on a pixel-by-pixel basis, and theoretically should allow us to determine whether the emission observed at a specific pixel is generated within or behind the turbulent volume. This is an advantage over the structure function method developed by \cite{Sun2014}, which involves calculating an average over a portion of the produced image, as our methods provide local information about the observed turbulent region.

\section{Method to Map the Rotation Measure}
\label{map_Fara}

In Paper I, we postulated that the polarization wavelength derivative and wavelength curvature could provide a rotationally and translationally invariant method of determining the rotation measure, by avoiding analysis of the polarization angle. For example, for the case of backlit emission, the wavelength derivative is the same as the rotation measure. This may provide more information on the underlying turbulence, such as the fluctuations in the electron density, and the structure of the Galactic magnetic field. In this section, we investigate what information our polarization diagnostics provide on the rotation measure, for the case of internal emission.

\begin{figure*}
\begin{center}
\includegraphics[scale=0.80]{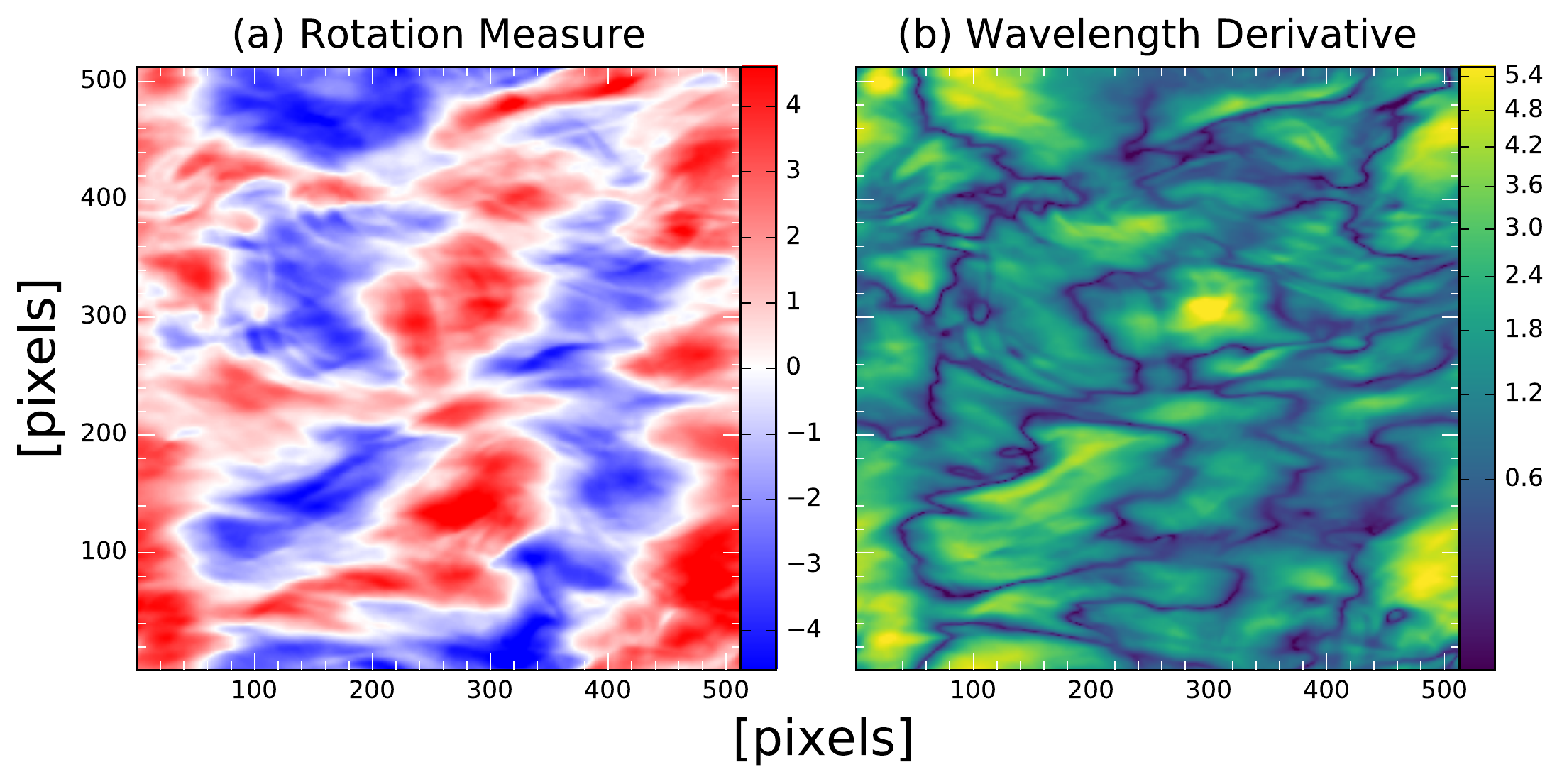}
\caption{The rotation measure (a, left) in units of rad m$^{-2}$, and wavelength derivative (b, right) in units of m$^{-2}$, for the Ms0.5Ma0.7 simulation, and a line of sight along the $z$ axis. The image of the wavelength derivative is produced for internal emission, at a frequency of $1.58$ GHz.}
\label{Fara_wav_grad}
\end{center}
\end{figure*}

In Fig. \ref{Fara_wav_grad} we show an example image of the rotation measure (left) and wavelength derivative (right), for the Ms0.5Ma0.7 simulation, internal emission, and a line of sight perpendicular to the mean magnetic field, at a frequency of $1.58$ GHz. This frequency corresponds to the third wavelength slice of the data cube, which was chosen because the first two and last two slices of the data cube suffer from numerical errors caused by the method of calculating second order derivatives. We find that areas where the wavelength derivative is zero are very well correlated with areas where the rotation measure is zero, and also that the wavelength derivative tends to attain large values in areas where the magnitude of the rotation measure is large. This occurs for all sub-\alf ic simulations, provided that the line of sight is perpendicular to the mean magnetic field. If the line of sight is parallel to the mean magnetic field, then the wavelength derivative resembles the polarization intensity. If the mean magnetic field is weak (super-\alf ic), then the wavelength derivative resembles the rotation measure, modulated by the polarization intensity.

We refer to regions where the wavelength derivative is zero as `depolarization interference fringes' (the black filaments in Fig. \ref{Fara_wav_grad}). There are three possible causes for these fringes:
\begin{enumerate}
\item The polarization intensity is zero along the fringe, at this wavelength.
\item The rotation measure is zero along the fringe.
\item The superposition of polarization vectors along the line of sight is such that the observed polarization vector does not depend on wavelength, at this wavelength.
\end{enumerate}
It is possible to determine which fringes are caused by the polarization intensity being zero by comparing the wavelength derivative to the polarization intensity. If we only examine fringes that are not seen in polarization intensity, then those that change with wavelength must be caused by a superposition of polarization vectors that happen to have no wavelength dependence at a single wavelength, and those that do not change with wavelength must be places where the rotation measure is zero.

In addition to the wavelength derivative being large in places of high rotation measure, we observe that depolarization interference fringes appear to emanate away from local maxima and minima of rotation measure as the observing wavelength increases. This can help us to pinpoint these local maxima and minima of the rotation measure, and obtain an idea of what the contours of the rotation measure look like around these positions, provided that the line of sight is perpendicular to the mean magnetic field.  

\begin{figure*}
\begin{center}
\includegraphics[scale=0.70]{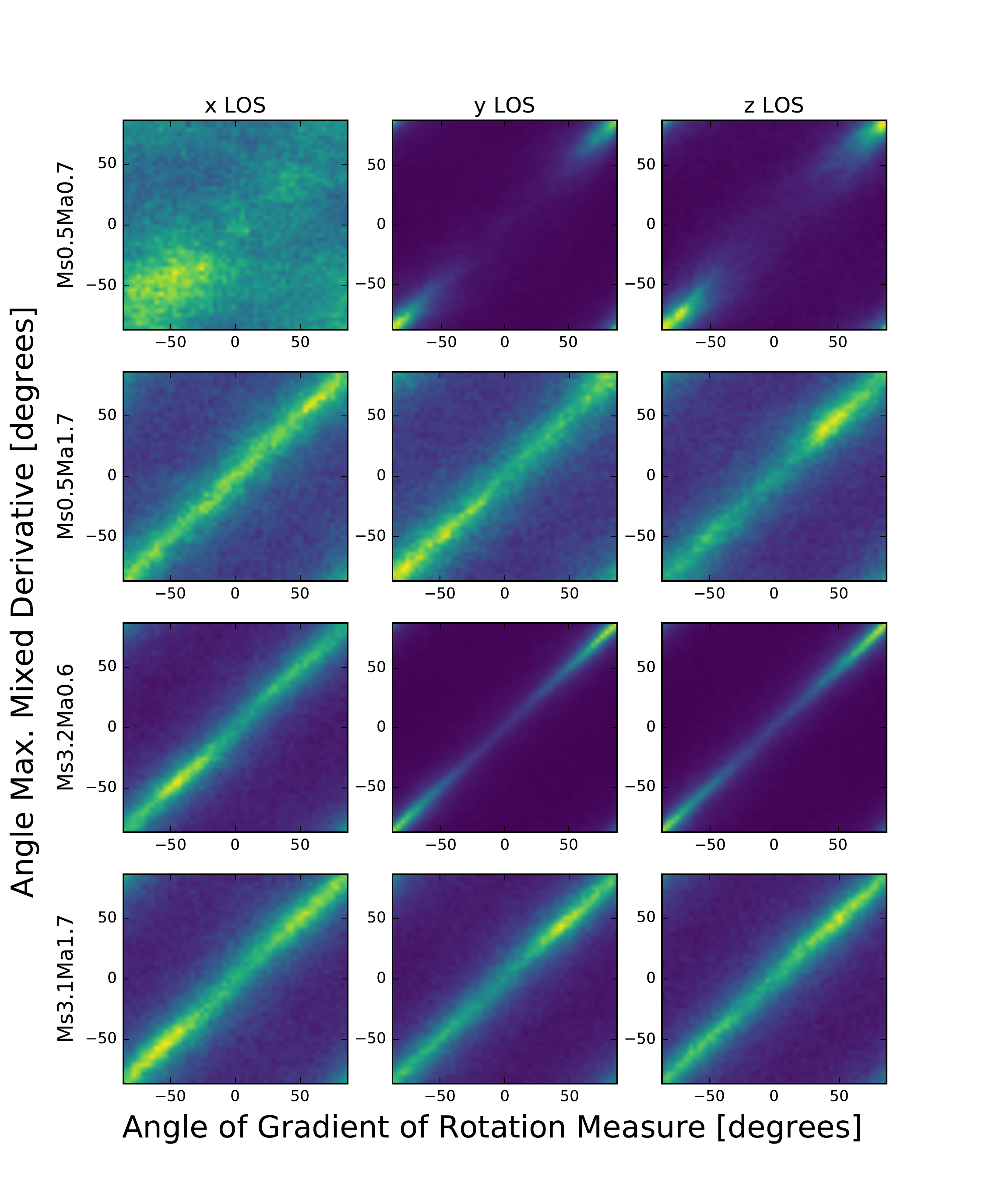}
\caption{Correlation plots of the angle that maximizes the mixed derivative ($y$ axis of each plot, in degrees) against the angle of the gradient of the rotation measure ($x$ axis of each plot, in degrees), for lines of sight along the $x$ (left column), $y$ (middle column), and $z$ (right column) axes. The top row is for the Ms0.5Ma0.7 simulation, the second row for Ms0.5Ma1.7, the third row for Ms3.2Ma0.6, and the bottom row for Ms3.1Ma1.7, all for the case of internal emission, at a frequency of $1.58$ GHz. Each correlation plot is a scatter plot of the values in the corresponding images, viewed as a heatmap, such that yellow represents the maximum number density of points in the scatter plot, and black represents a number density of zero.}
\label{corr_plots_angles}
\end{center}
\end{figure*}

We find an excellent degree of correlation between the angle that maximizes the polarization mixed derivative, and the angle of the gradient of the rotation measure, for all of our simulations, and almost all lines of sight. We demonstrate this correlation in Fig. \ref{corr_plots_angles}, which shows scatter plots of values of the angle that maximizes the mixed derivative against the corresponding values of the angle of the gradient of the rotation measure. These scatter plots are shown as heatmaps, such that yellow represents a large number of points in that area of the scatter plot. Lines of sight along the $x$ (parallel to the mean magnetic field), $y$, and $z$ axes are shown in the left, middle, and right columns respectively, and from top to bottom, the rows give the scatter plots for the Ms0.5Ma0.7, Ms0.5Ma1.7, Ms3.2Ma0.6, and Ms3.1Ma1.7 simulations, at a frequency of $1.58$ GHz, for internal emission. 

With the exception of the Ms0.5Ma0.7 simulation and a line of sight along the $x$ axis, we find clear linear relationships between the angle that maximizes the mixed derivative, and the angle of the gradient of the rotation measure, at this wavelength. At long wavelengths, namely at a frequency of $0.5$ GHz, we find approximately linear relationships between these variables for all simulations and lines of sight, although for supersonic simulations (bottom two rows), the correlation is not as tight. The angle that maximizes the mixed derivative is hence an excellent tracer of the angle of the gradient of the rotation measure, as correlation plots such as those shown in Fig. \ref{corr_plots_angles} can be used to determine the angle of the gradient of the rotation measure to an accuracy of approximately $10^{\circ}$, for most regimes of turbulence and lines of sight.

By combining the wavelength derivative and the angle that maximizes the mixed derivative, it is possible to determine the locations of maximum, minimum, and zero rotation measure, as well as the angle of the gradient of the rotation measure, from which the contours of the rotation measure can be determined. This provides us with a good idea of what the underlying rotation measure looks like. If it becomes possible to image the gradient of the rotation measure in the future, then we will be able to produce images of the rotation measure itself.

\section{Discussion}
\label{discuss}

We have found that images of the polarization gradient calculated for the case of internal emission look similar to observed polarization gradients in the CGPS \citep{Herron2017b}. As a result, it seems plausible that many polarimetric observations are of internally generated emission, and so any statistical method to constrain properties of turbulence from polarimetric observations, similar to the methods proposed by \cite{Burkhart2012}, must first determine whether the emission is backlit or internal. The methods of distinguishing between backlit and internal emission on a pixel-by-pixel basis that we have outlined in Section \ref{back_int} hence play an important role in the measurement of properties of turbulence, although further work is required to confirm that these methods are robust. It is also necessary for future work to extend the analysis of \cite{Burkhart2012} to the case of internal emission, for the polarization diagnostics presented in Paper I. 

Further work is required to confirm whether the maximum amplitudes of the radial and tangential components of the directional derivative can be used to qualitatively determine whether polarization gradient structures are caused by small-scale or large-scale fluctuations. With our current simulations we are only able to investigate small-scale fluctuations caused by turbulence, but simulations that have a gradient in the mean magnetic field (either in its strength or its direction), or a gradient of the depth along the line of sight, may be better suited to studying large-scale changes. Such simulations may provide insight on whether the radial and tangential components of the directional derivative can be used to investigate the relative importance of small- and large-scale changes in the magnetoionic medium on the observed polarization structures.

Throughout this work we have ignored beam depolarization; the destructive interference of polarization vectors within the telescope beam. This effect would lower the polarization intensity measured in our synthetic observations, and would also introduce spatial and spectral dependence into the polarization intensity for the case of backlit emission. Including beam depolarization may have a strong impact on our proposed methods to distinguish backlit and internal emission using spatial derivatives of polarization, such as the polarization directional curvature, as it will cause the observed polarization intensity to be non-uniform in the case of backlit emission, and the directional curvature to not be equal to $1/P$. The methods to distinguish between backlit and internal emission using spectral diagnostics should not be as strongly affected, as it is possible to smooth images produced at different observing frequencies such that the angular resolution is the same for all images. This would help to counteract the spectral dependence that beam depolarization introduces into synthetic observations of backlit emission, due to the changing shape of the telescope beam. However, this does not negate the spectral dependence entirely, as the Faraday rotation of polarization vectors will differ for vectors within the beam, so that the degree of destructive interference varies with wavelength. 

Beam depolarization will also have an effect on our preliminary method of mapping features of the rotation measure, from which we can study the structure of the Galactic magnetic field, as there may not be any wavelength-independent depolarization interference fringes when beam depolarization is included. Additionally, beam depolarization may affect the correlation between the angle that maximizes the mixed derivative, and the angle of the gradient of the rotation measure. Future work examining the effect of beam depolarization on synthetic images of our polarization diagnostics will be required to ensure that our methods for distinguishing between backlit and internal emission, and for mapping the rotation measure, are robust. It is also important to examine the influence of noise on these methods, particularly the maximum amplitudes of the radial and tangential components of the directional derivative, as these diagnostics are not translationally invariant.

From our qualitative analysis, we have found systematic changes in the observed structures of the polarization diagnostics that could be used to constrain properties of turbulence. These findings are described in detail in Appendix \ref{depend}, and we briefly summarize them here. A common finding for our polarization diagnostics is that structures tend to be elongated for lines of sight perpendicular to a strong magnetic field, and tend to have more small scale structure for lines of sight parallel to a strong field. Lines of sight parallel to the mean magnetic field can also be more wavelength dependent than other lines of sight, as they have a larger rotation measure than lines of sight perpendicular to the mean magnetic field. 

Similar to \cite{Burkhart2012}, we also find that supersonic simulations tend to have more small scale structure than subsonic simulations, and sub-Alfv\'enic simulations have more elongated structures than super-Alfv\'enic structures, for lines of sight perpendicular to the mean magnetic field. These findings imply that we could use statistics that quantify how strong fluctuations are on small-scales, such as the slope of a structure function, or how elongated they are, such as the quadrupole ratio (see \citealt{Herron2016} for more information), to constrain the sonic and Alfv\'enic Mach numbers, and the direction of the mean magnetic field. We caution that as the observed structures are wavelength dependent, the relationship between statistics of polarization diagnostics and properties of the turbulence may also be sensitive to wavelength. Future work is required to quantify how statistics of these diagnostics are related to properties of turbulence, and how these relationships change with wavelength. Such work will complement the research conducted on the methods proposed by \cite{Lazarian2016}.

Other promising statistics include the Minkowski functionals (\citealt{Minkowski1903}, see \citealt{Mecke1994, Schmalzing1997} for more information), which are a complete set of morphological descriptors, that can be calculated for a surface defined by a specified isodensity contour. For a two-dimensional region, the Minkowski functionals are the circumference, enclosed area, and genus\footnote{See \url{https://golem.ph.utexas.edu/category/2011/06/hadwigers\_theorem\_part\_1.html} for more information.}, and for a three-dimensional surface, the Minkowski functionals include the volume enclosed by the surface, its surface area, its integrated mean curvature, and the integrated Gaussian curvature (which is related to the Euler characteristic and genus, see \citealt{Mecke1994} for more information on Minkowski functionals in three dimensions). 

It is possible to calculate the Minkowski functionals of two-dimensional regions for images of the polarization diagnostics derived in Paper I, and these statistics may provide robust constraints on the properties of the turbulence, similar to the finding by \cite{Burkhart2012} that the genus of the polarization gradient is sensitive to the sonic Mach number. It is also possible to consider the polarization diagnostics in three-dimensions, with wavelength as the third axis, and then Minkowski functionals can be calculated for three-dimensional structures defined in this cube. 

\section{Conclusions and Future Work}
\label{concl}

We have generated synthetic maps of Stokes $Q$ and $U$ for simulations of ideal magnetohydrodynamic turbulence, for the cases where the turbulent volume is illuminated from behind by polarized emission, and where the emission comes from within the volume. Using these synthetic maps, we have calculated all of the invariant polarization diagnostics derived in Paper I for each simulation, and different lines of sight, between frequencies of $0.5$ and $2$ GHz.

We have found that the polarization gradient and generalized polarization gradient trace spatial changes in the magnetoionic medium for the case of internal emission, provided that depolarization is not severe. Images of the polarization gradient for supersonic simulations, and internal emission, display similar features to those observed in the Canadian Galactic Plane Survey at low longitudes, and so this region may be supersonic. This also suggests that a significant fraction of observed polarized emission is generated within turbulent regions, and so it is necessary to determine whether we observe backlit or internal emission before attempting to constrain properties of turbulence statistically.

We have detailed methods that could be used to distinguish between backlit and internal emission, using the polarization directional curvature and the polarization wavelength curvature. These methods work on a pixel-by-pixel basis, however assume perfect angular resolution, and so it will be necessary to study how robust these methods are for finite angular resolution.

We have discussed a preliminary method that could be used to create maps of the rotation measure, which would provide information on the structure of the Galactic magnetic field. This method involves using the polarization wavelength derivative to determine where the rotation measure is zero, or attains local maximum or minimum values, and using the angle that maximizes the mixed derivative to determine the direction of the gradient of the rotation measure. From this information, it is possible to reconstruct the contours of the rotation measure.

For the polarization diagnostics we have examined, we found that supersonic simulations tend to have more small-scale structure than subsonic simulations, and lines of sight parallel to the mean magnetic field have more small-scale structure than lines of sight perpendicular to the mean magnetic field. Features of these diagnostics tend to be elongated along the mean magnetic field, provided the perpendicular component of the magnetic field is strong, and the degree of elongation is greater for lower Alfv\'enic Mach number. We speculate that statistics of these diagnostics, such as the Minkowski functionals, could be used to provide constraints on the sonic and Alfv\'enic Mach numbers, and the direction of the mean magnetic field. These statistics will depend on the observing wavelength, however, and this must be taken into consideration.

\section*{Acknowledgements}

We thank Xiaohui Sun for providing the maps of Stokes $Q$ and $U$ used to produce Fig. \ref{Xiaohui_curv}. C.~A.~H. acknowledges financial support received via an Australian Postgraduate Award, and a Vice Chancellor's Research Scholarship awarded by the University of Sydney. B.~B. is supported by the NASA Einstein Postdoctoral Fellowship. B.~M.~G. acknowledges the support of the Natural Sciences and Engineering Research Council of Canada (NSERC) through grant RGPIN-2015-05948 and of a Canada Research Chair. N.~M.~M.-G. acknowledges the support of the Australian Research Council through grant FT150100024. The Dunlap Institute for Astronomy and Astrophysics is funded through an endowment established by the David Dunlap family and the University of Toronto. This work has been carried out in the framework of the S-band Polarisation All Sky Survey (S-PASS) collaboration. The Parkes Radio Telescope is part of the Australia Telescope National Facility, which is funded by the Commonwealth of Australia for operation as a National Facility managed by CSIRO. This research made use of Astropy, a community-developed core Python package for Astronomy \citep{Astropy2013}, and APLpy, an open-source plotting package for Python \citep{Robitaille2012}.

\appendix
\section{Appendix A: Dependence of Diagnostics on Line of Sight, Wavelength, and Regime of Turbulence}
\label{depend}

In this appendix we will discuss how the polarization diagnostics derived in Paper I depend on the line of sight and wavelength used to produce the synthetic images of Stokes $Q$ and $U$, and on the regime of turbulence of the simulations. For all diagnostics, we find that there is no dependence on the line of sight for simulations with a weak magnetic field (super-Alfv\'enic), and so we will only discuss line of sight dependence for simulations with a strong magnetic field. We will also only discuss the case of internal emission, unless otherwise stated.

\subsection{A.1 First Order Spatial Derivatives} \label{first_space}
\begin{figure*}
\begin{center}
\includegraphics[scale=0.58]{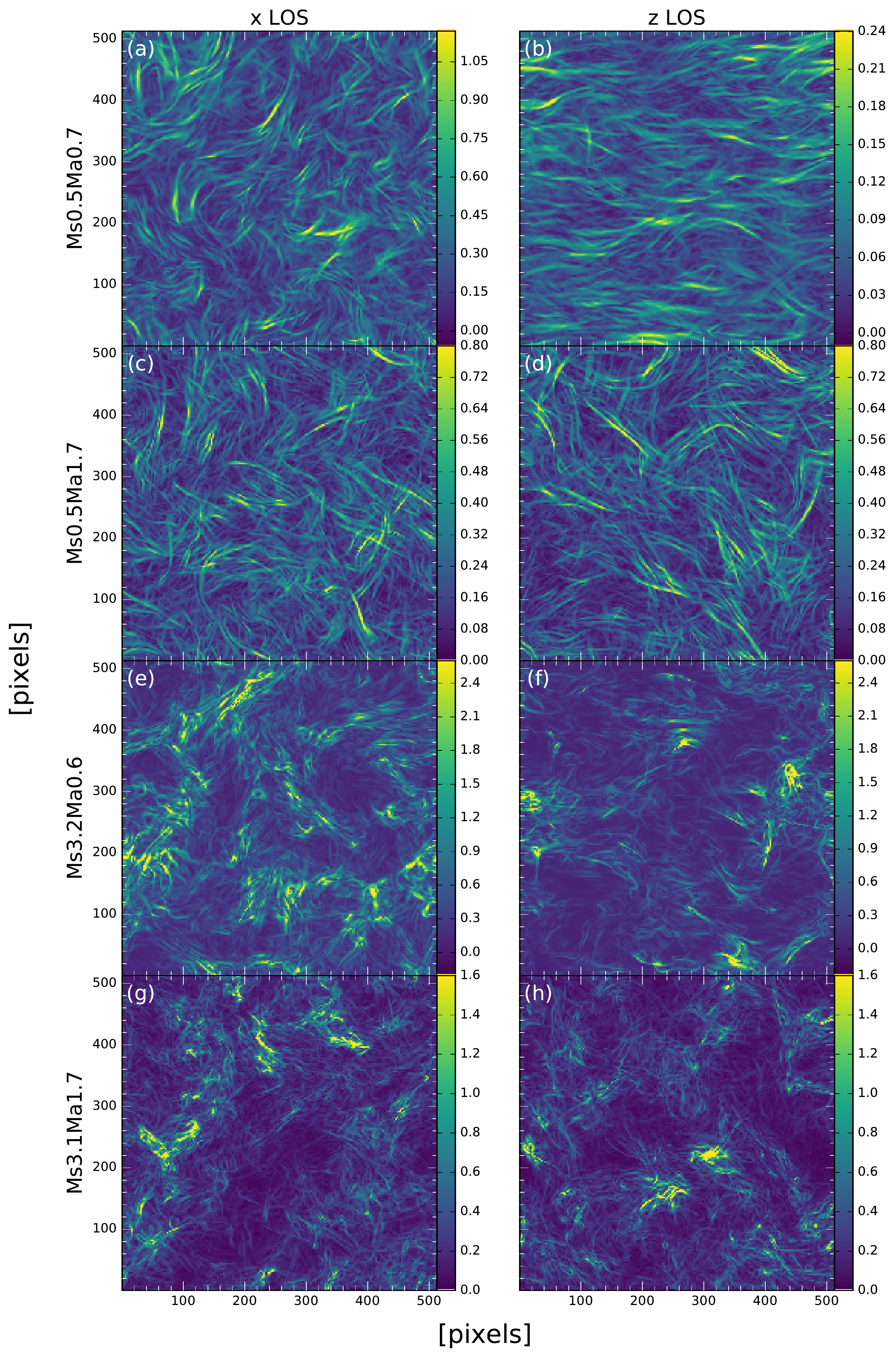}
\caption{The generalized polarization gradient for the Ms0.5Ma0.7 (top row), Ms0.5Ma1.7 (second row), Ms3.2Ma0.6 (third row), Ms3.1Ma1.7 (bottom row) simulations, and lines of sight along the $x$ (left column) and $z$ (right column) axes. All images were produced for internal emission, and a frequency of $2$ GHz. Units are pc$^{-1}$. Different color scalings are used for the images.}
\label{max_direc_deriv_short}
\end{center}
\end{figure*}

\begin{figure*}
\begin{center}
\includegraphics[scale=0.58]{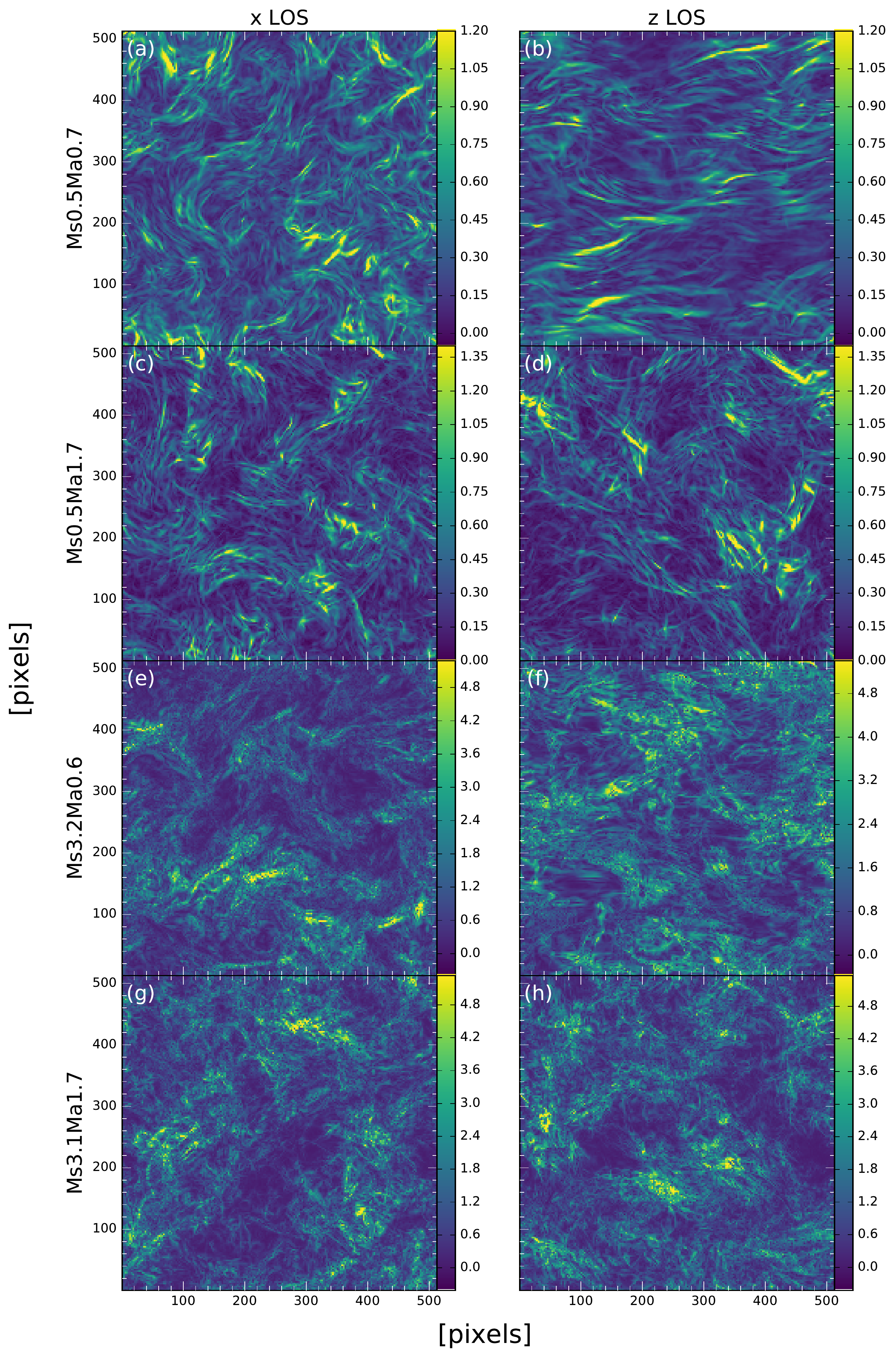}
\caption{The same as Fig. \ref{max_direc_deriv_short}, but for a frequency of $0.5$ GHz. Different color scalings are used for the images.}
\label{max_direc_deriv_long}
\end{center}
\end{figure*}

In this section we discuss the generalized polarization gradient, the angle that maximizes the directional derivative, the maximum amplitudes of the radial and tangential components of the directional derivative, and the angles that maximize the radial and tangential components of the directional derivative. In Fig. \ref{max_direc_deriv_short} we show images of the generalized polarization gradient for the Ms0.5Ma0.7 (top row), Ms0.5Ma1.7 (second row), Ms3.2Ma0.6 (third row), Ms3.1Ma1.7 (bottom row) simulations, and lines of sight along the $x$ (left column) and $z$ (right column) axes, for internal emission, and a frequency of $2$ GHz. In Fig. \ref{max_direc_deriv_long} we show the corresponding images of the generalized polarization gradient at a frequency of $0.5$ GHz.

As the first order spatial derivatives are related to the polarization directional derivative, they exhibit similar dependencies on the line of sight, wavelength, and regime of turbulence, in general. We find that for simulations with a strong magnetic field (sub-Alfv\'enic), there can be differences between lines of sight that are parallel or perpendicular to the mean magnetic field. If the simulation is subsonic, then lines of sight perpendicular to the mean field will show features that are elongated in the direction of the magnetic field. If the simulation is supersonic, then different lines of sight look fairly similar at short wavelengths (there is little elongation for lines of sight perpendicular to a strong magnetic field), but different at long wavelengths, because of enhanced depolarization along the line of sight parallel to the mean magnetic field. There is also an increasing degree of small-scale structure as the wavelength increases.

For subsonic simulations, the observed structures have little dependence on wavelength in general, but the generalized polarization gradient is an exception to this. For supersonic simulations, clear structures observed at short wavelengths are slowly replaced by a small-scale depolarisation pattern that appears to be superimposed over a larger-scale pattern. For the generalized polarization gradient, the small-scale depolarization pattern appears to grow outwards from the bright regions seen at short wavelengths.

We have also found that:
\begin{itemize}
\item The angle that maximizes the directional derivative is the same as the angle of the gradient of the rotation measure, for backlit emission. For internal emission, there is a weak correlation between the angle that maximizes the directional derivative, and the angle of the gradients of the rotation measure, but only at short wavelengths.

\item The maximum amplitude of the radial component of the directional derivative has significant wavelength dependence for supersonic simulations, which is strongest for lines of sight perpendicular to the field.

\item The maximum amplitude of the tangential component of the directional derivative has more small-scale structure for lines of sight parallel to a strong magnetic field, than perpendicular to the field. For subsonic simulations, the brightness of the maximum amplitude of the tangential component increases with wavelength, because the increasing amount of Faraday rotation and depolarisation can lead to larger changes in the polarisation angle. The contrast between bright and faint filaments also increases with wavelength, which can be partly attributed to depolarisation.
\end{itemize}

\subsection{A.2 Second Order Spatial Derivatives} \label{second_space}

\begin{figure*}
\begin{center}
\includegraphics[scale=0.58]{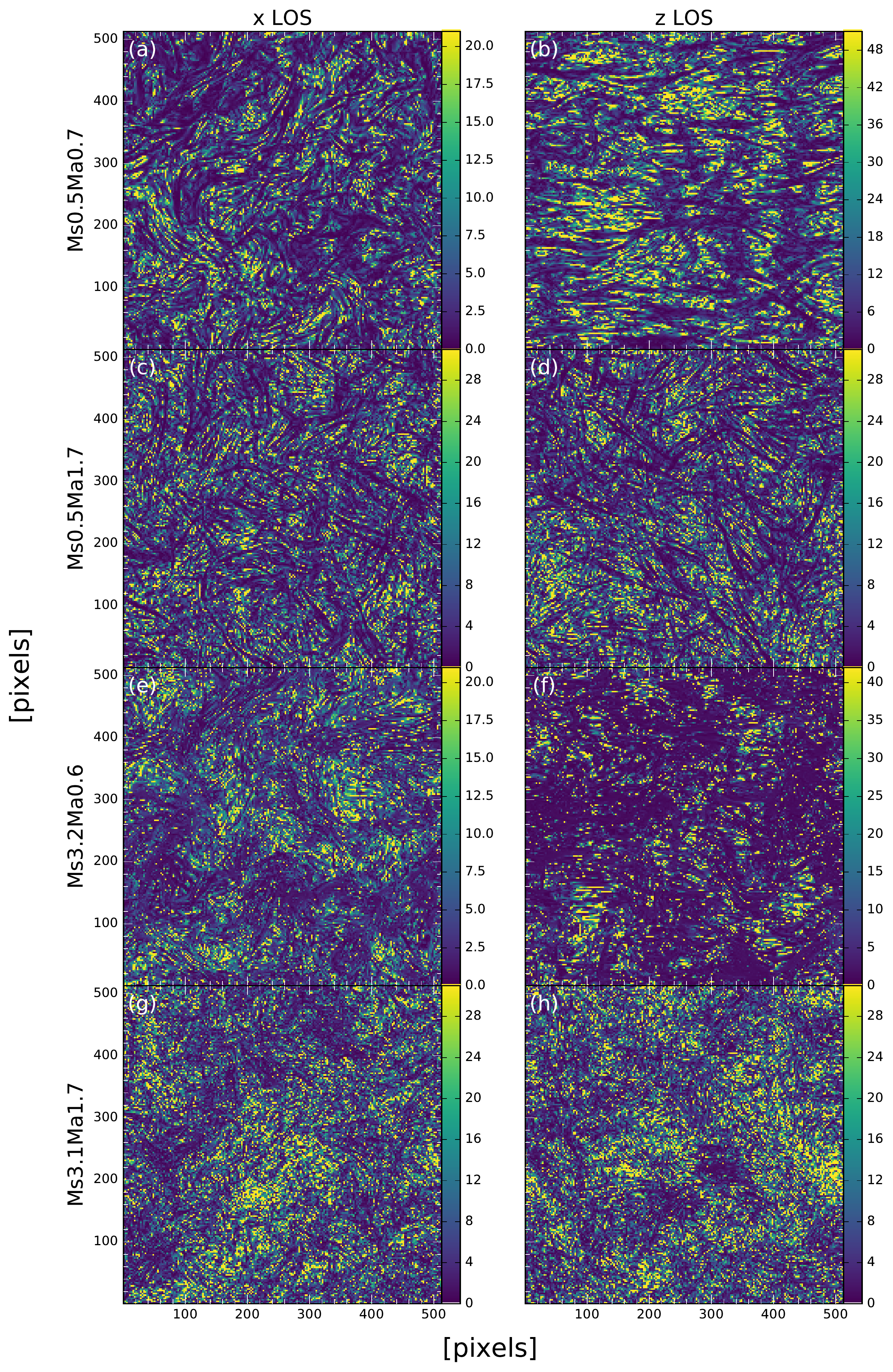}
\caption{The same as Fig. \ref{max_direc_deriv_short}, but for the polarization directional curvature in the direction that maximizes the directional derivative, instead of the generalized polarization gradient. Units are pc$^{-2}$. Different color scalings are used for the images.}
\label{curv_max_direc_short}
\end{center}
\end{figure*}

\begin{figure*}
\begin{center}
\includegraphics[scale=0.58]{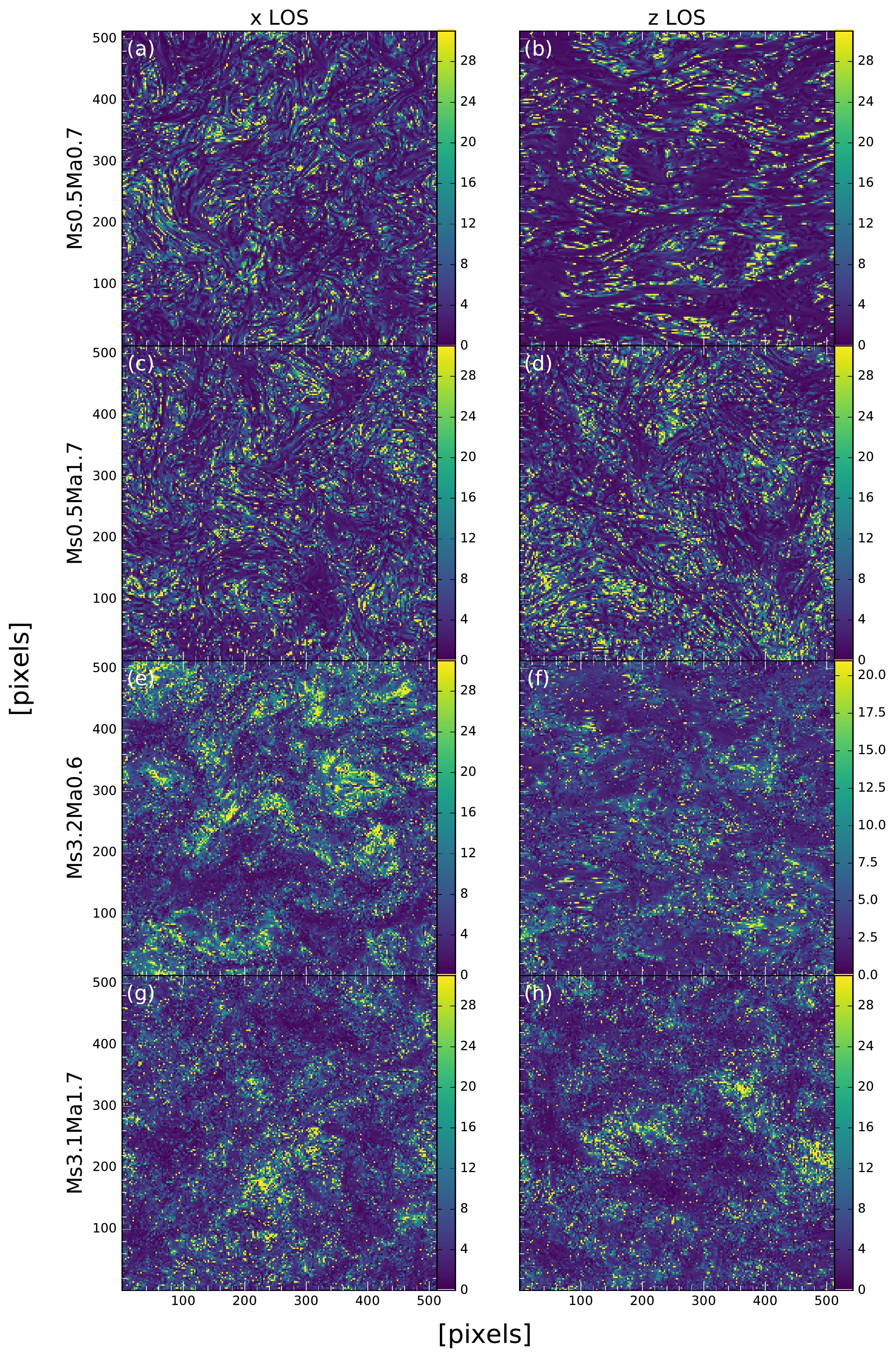}
\caption{The same as Fig. \ref{curv_max_direc_short}, but for a frequency of $0.5$ GHz. Different color scalings are used for the images.}
\label{curv_max_direc_long}
\end{center}
\end{figure*}

The polarization directional curvature in the direction that maximizes the directional derivative is derived from the polarization directional curvature, and so the two have very similar dependencies on the line of sight, wavelength, and regime of turbulence. In Fig. \ref{curv_max_direc_short} we show images of the polarization directional curvature in the direction that maximizes the directional derivative, for the same simulations, lines of sight, and frequency as Fig. \ref{max_direc_deriv_short}. Fig. \ref{curv_max_direc_long} shows the corresponding directional curvature images at a frequency of $0.5$ GHz.

For internal emission, large polarization directional curvature tends to correspond to maxima and minima of the polarisation intensity, or to maxima and minima of the polarisation angle, provided that the rate of change of the other polarisation quantity is large. For example, the curvature will be large at a maximum of polarization intensity, if the rate of change of polarization angle is large. 

For lines of sight parallel to a strong magnetic field, the polarization directional curvature shows filaments that are evenly spaced, but the spacing between filaments varies for lines of sight perpendicular to a strong field. Lines of sight perpendicular to a strong magnetic field also tend to have filaments aligned with the field, if the directional curvature is not calculated in a direction parallel to the field. Lines of sight perpendicular to the magnetic field are more sensitive to wavelength than lines of sight parallel to the field, and curvature values are larger for lines of sight perpendicular to a strong magnetic field at short wavelengths than for lines of sight parallel to the magnetic field, but smaller for these lines of sight at long wavelengths. 

We find that supersonic simulations have more structure on small scales than subsonic simulations, and the amount of small-scale structure increases with wavelength. The magnitude of the curvature also appears to decrease with increasing wavelength. For subsonic simulations, only those with a weak magnetic field are sensitive to wavelength, if the line of sight is perpendicular to the field. For all simulations, the curvature can be large in regions of low polarisation intensity.

\subsection{A.3 First Order Wavelength Derivatives} \label{first_wave}

\begin{figure*}
\begin{center}
\includegraphics[scale=0.58]{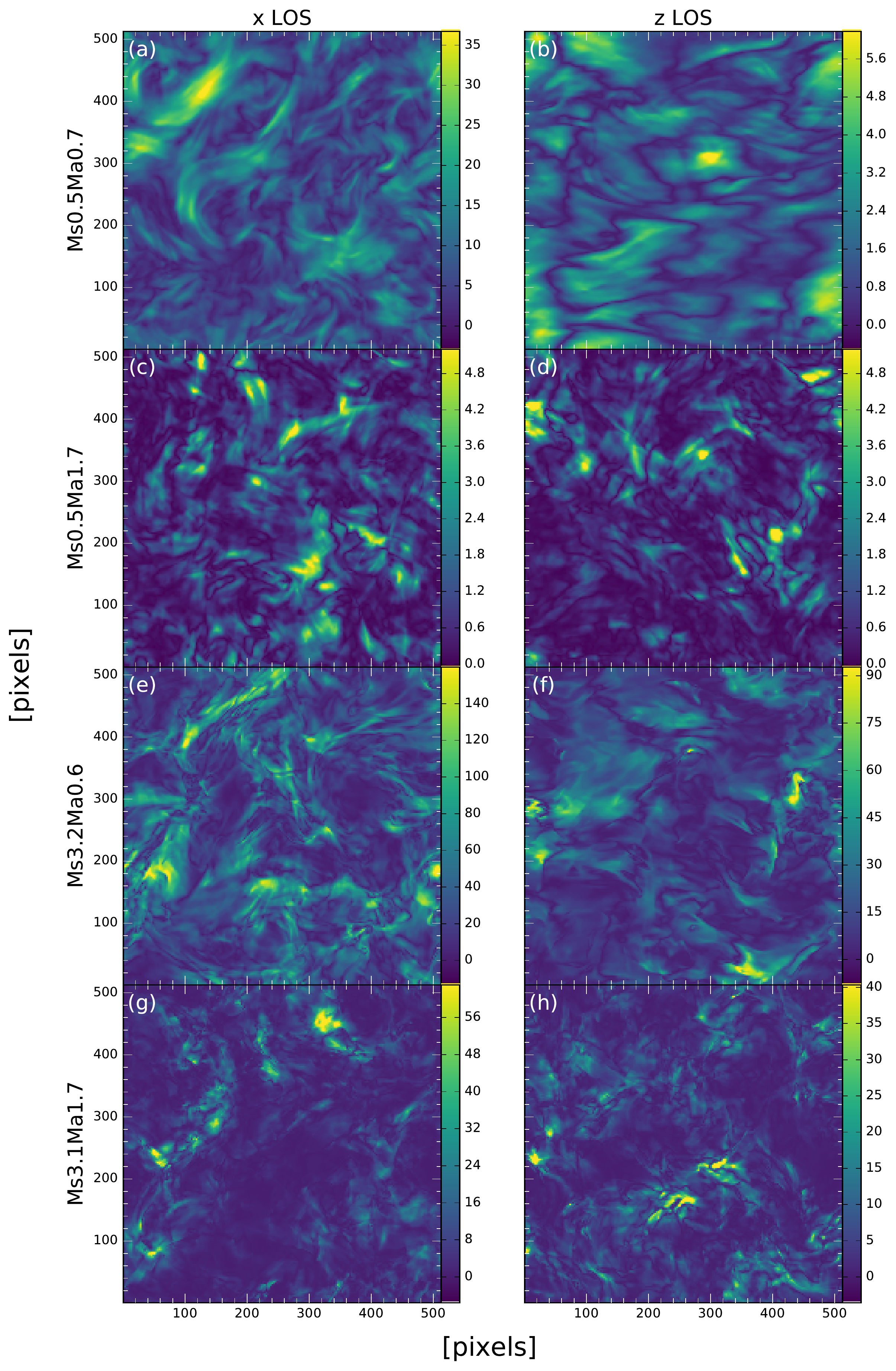}
\caption{The same as Fig. \ref{max_direc_deriv_short}, but for the polarization wavelength derivative, instead of the generalized polarization gradient, at a frequency of $1.58$ GHz. Units are m$^{-2}$. Different color scalings are used for the images.}
\label{wav_grad_short}
\end{center}
\end{figure*}

\begin{figure*}
\begin{center}
\includegraphics[scale=0.58]{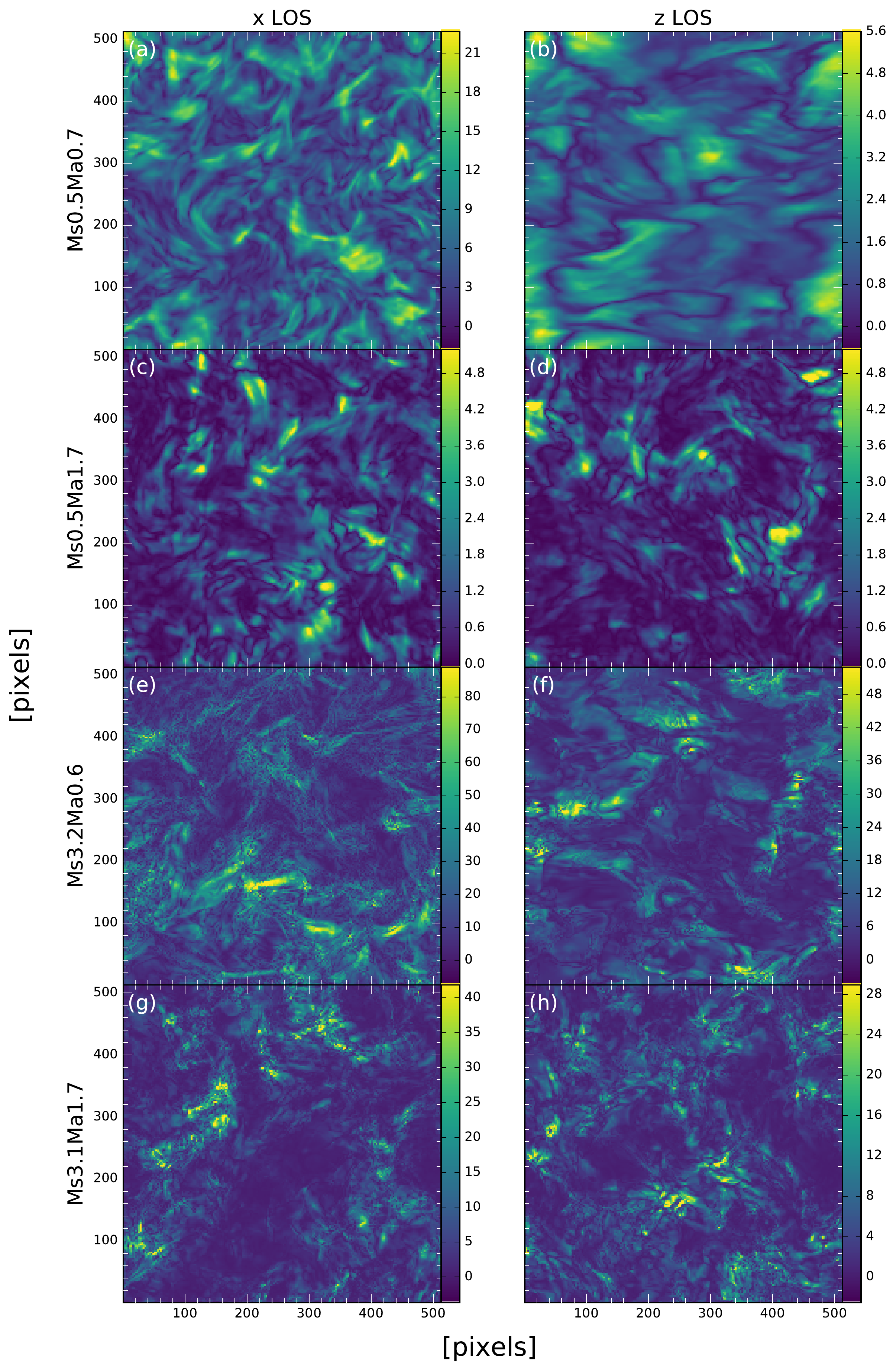}
\caption{The same as Fig. \ref{wav_grad_short}, but for a frequency of $0.51$ GHz. Different color scalings are used for the images.}
\label{wav_grad_long}
\end{center}
\end{figure*}

For internal emission, the polarization wavelength derivative is related to the first order wavelength derivatives of the polarization intensity and the polarization angle, where the latter is weighted by polarization intensity. In Fig. \ref{wav_grad_short}, we show the polarization wavelength derivative for the same simulations and lines of sight as Fig. \ref{max_direc_deriv_short}, at a frequency of $1.58$ GHz. In Fig. \ref{wav_grad_long} we show the corresponding images for a frequency of $0.51$ GHz.

For lines of sight perpendicular to a strong magnetic field, the wavelength derivative is large in areas of large rotation measure at short wavelengths, and small in areas where the rotation measure is zero. For lines of sight parallel to the mean magnetic field, the wavelength derivative largely traces polarization intensity, at all wavelengths. If the line of sight is perpendicular to a weak magnetic field, then the wavelength derivative tends to trace the rotation measure, modulated by the polarization intensity.

The wavelength derivative tends to decrease with increasing wavelength, due to depolarization, and there is an increasing amount of small-scale structure. If the line of sight is perpendicular to a strong magnetic field, the wavelength derivative is similar to the rotation measure at short wavelengths, and more like polarization intensity at long wavelengths. For simulations with a weak magnetic field, there is little dependence on wavelength.

\begin{figure*}
\begin{center}
\includegraphics[scale=0.58]{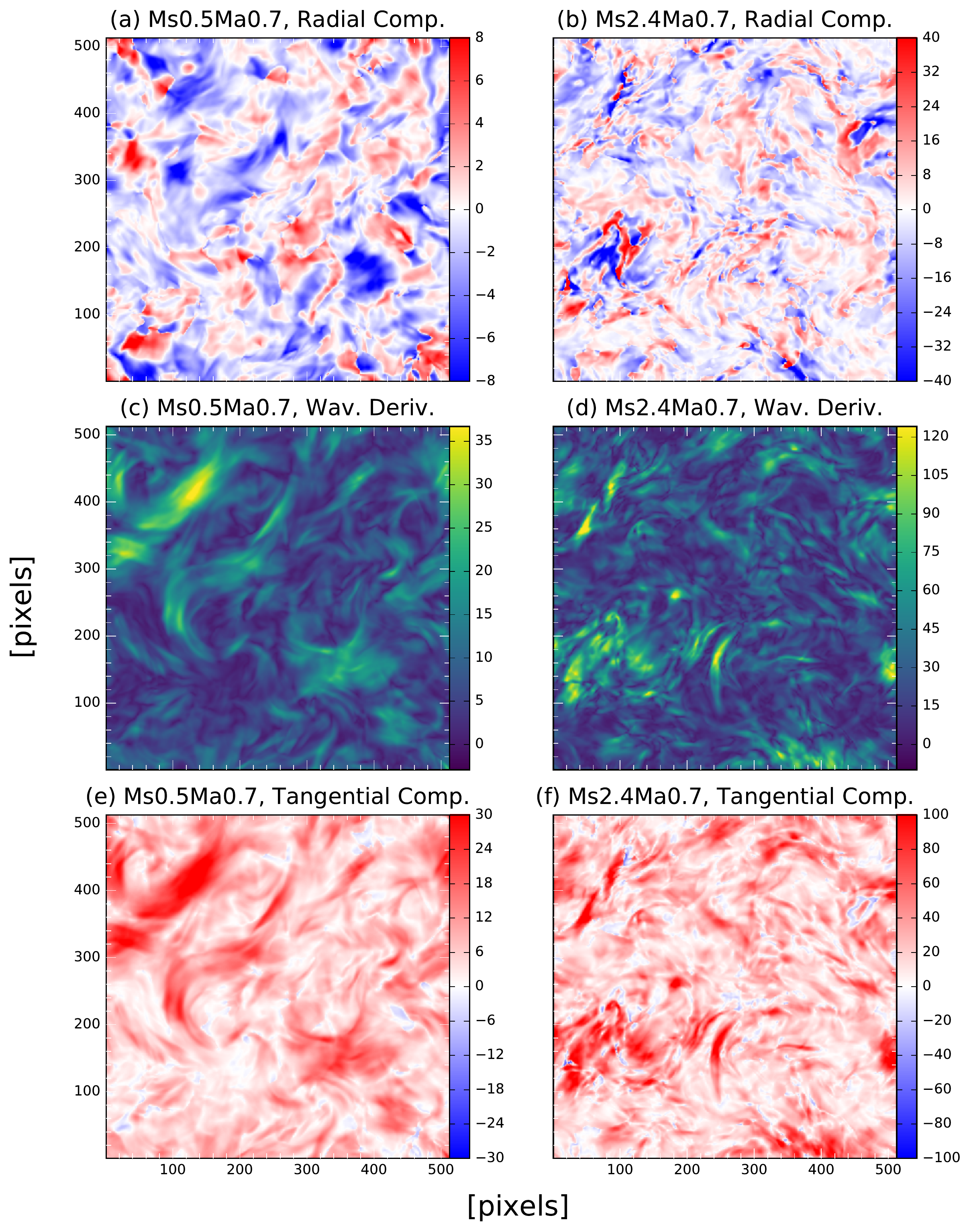}
\caption{A comparison of the radial (top row) and tangential (bottom row) components of the polarization wavelength derivative to the wavelength derivative (middle row), for the subsonic Ms0.5Ma0.7 simulation (left column) and supersonic Ms2.4Ma0.7 simulation (right column). These images were produced for internal emission, a line of sight along the $x$ axis, at a frequency of $1.58$ GHz. All images are in units of m$^{-2}$. Different color scalings are used for the images.}
\label{rad_tang_wav_grad}
\end{center}
\end{figure*}

In Fig. \ref{rad_tang_wav_grad} we show the polarization wavelength derivative (middle row), as well as the radial (top row) and tangential (bottom row) components of the wavelength derivative, for the Ms0.5Ma0.7 (left column) and Ms2.4Ma0.7 (right column) simulations. These images were produced for a line of sight parallel to the mean magnetic field, at a frequency of $1.58$ GHz. For these simulations, we find that the tangential component has more features in common with the wavelength derivative than the radial component. In general, we find that the tangential component always seems to be larger than the radial component for our simulations, which may be because the primary effect of Faraday rotation is to rotate the polarization angle.

We find that the radial component of the wavelength derivative is the same as the derivative of polarization intensity with respect to wavelength. For lines of sight perpendicular to a strong magnetic field, alternating positive and negative regions emanate from locations of high rotation measure, and this oscillation is more rapid for supersonic simulations. The observed structures tend to be aligned with the magnetic field for lines of sight perpendicular to the field. We observe that there is more small-scale structure for lines of sight parallel to a strong magnetic field, and that the amount of small-scale structure increases as the wavelength increases.

We observe that the tangential component of the wavelength derivative is the same as the rotation measure multiplied by polarization intensity. This allows us to image the rotation measure, without needing to worry about unwrapping the polarization angle, to account for situations where the polarization angle changes from close to $90^{\circ}$ to $-90^{\circ}$, or vice versa. The tangential component of the wavelength derivative is similar to the rotation measure for lines of sight perpendicular to the field, and similar to polarisation intensity for lines of sight parallel to the field. The observed structures tend to align with the magnetic field for lines of sight perpendicular to a strong field. As for the wavelength derivative, we find that more small-scale structure becomes apparent as the wavelength increases. 

\subsection{A.4 Second Order Wavelength Derivatives} \label{second_wave}

\begin{figure*}
\begin{center}
\includegraphics[scale=0.58]{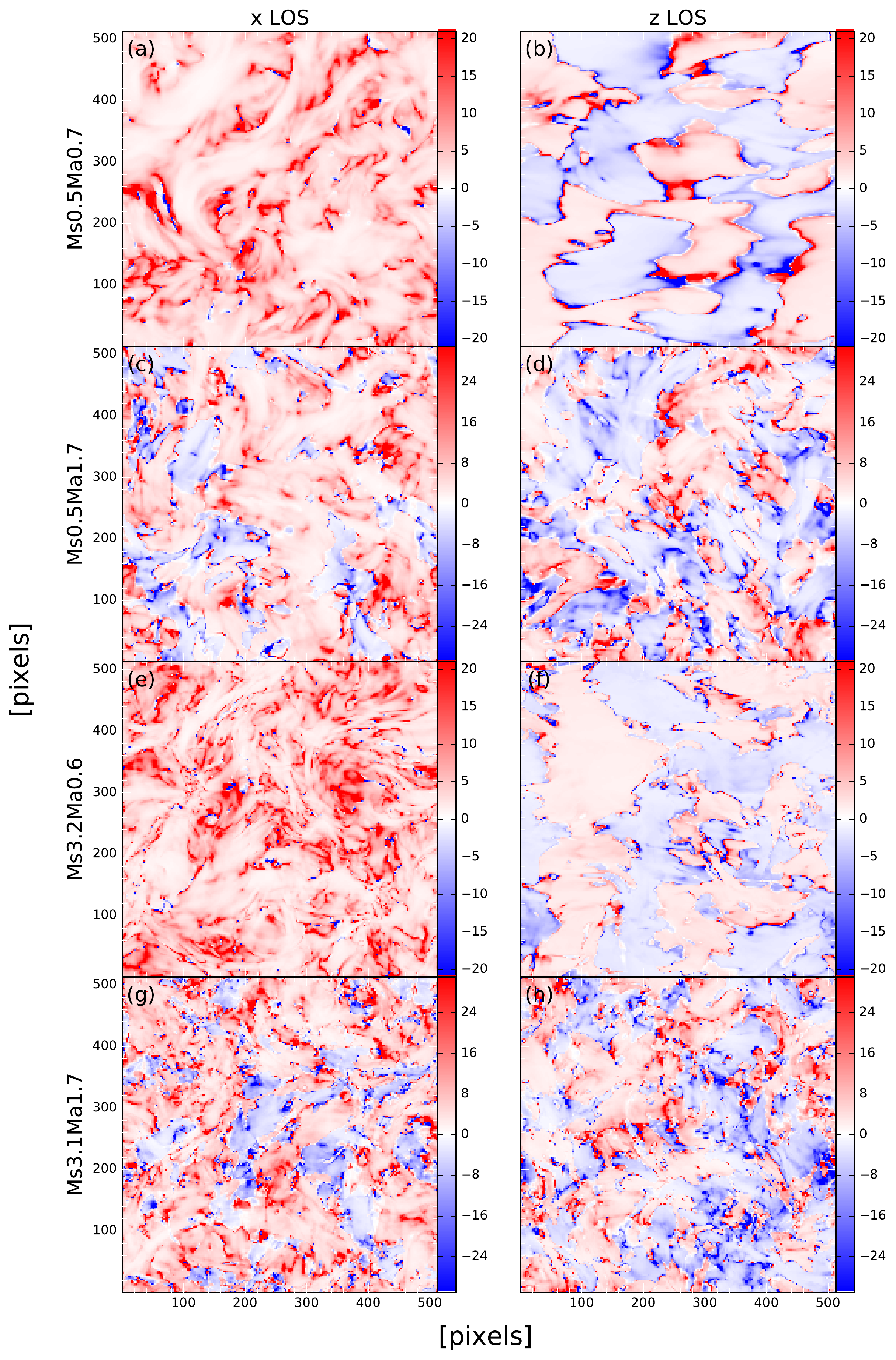}
\caption{The same as Fig. \ref{max_direc_deriv_short}, but for the polarization wavelength curvature, instead of the generalized polarization gradient, at a frequency of $1.58$ GHz. Units are m$^{-4}$. Different color scalings are used for the images.}
\label{wav_curv_short}
\end{center}
\end{figure*}

\begin{figure*}
\begin{center}
\includegraphics[scale=0.58]{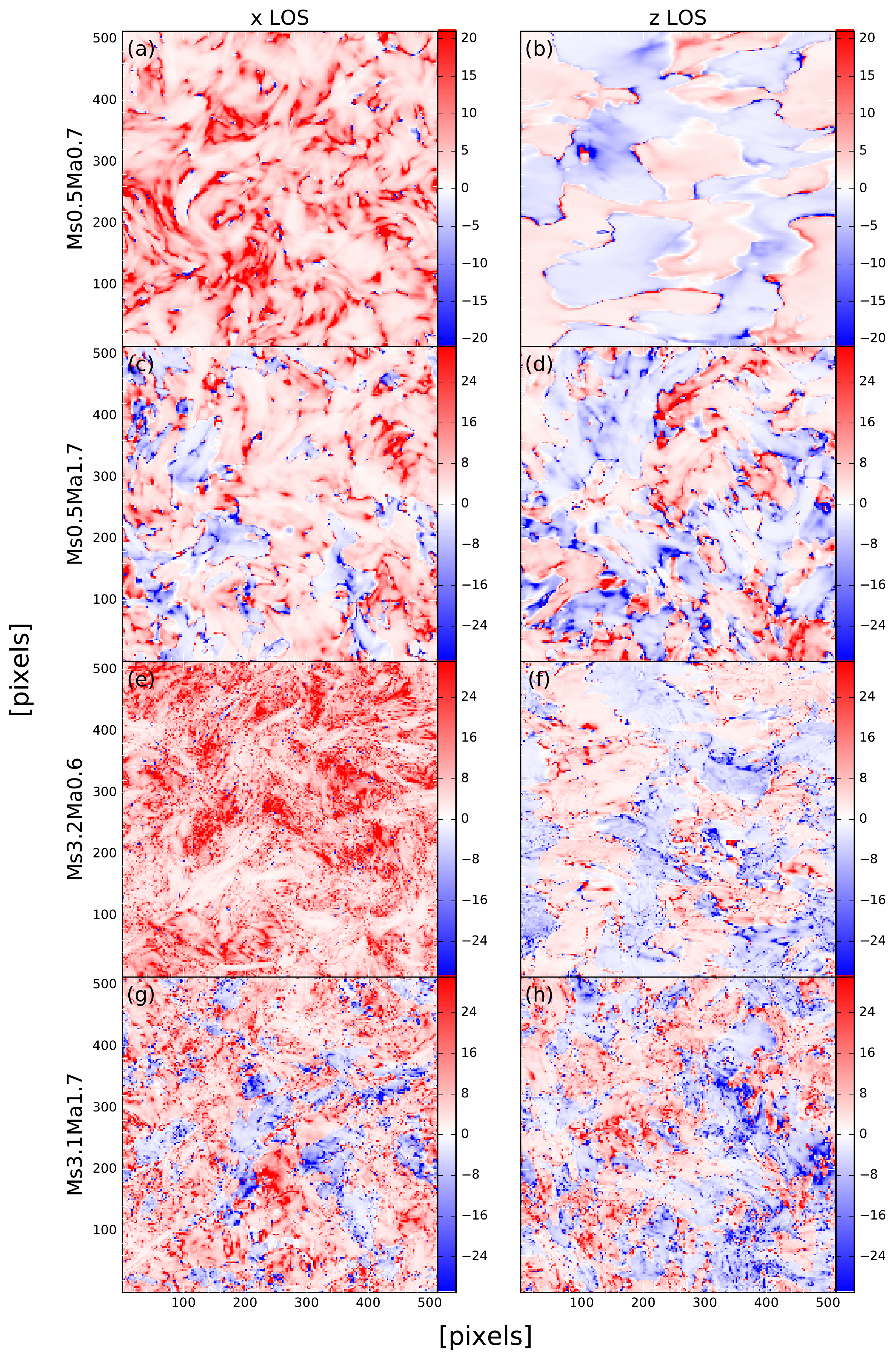}
\caption{The same as Fig. \ref{wav_curv_short}, but for a frequency of $0.51$ GHz. Different color scalings are used for the images.}
\label{wav_curv_long}
\end{center}
\end{figure*}

In Fig. \ref{wav_curv_short}, we show the polarization wavelength curvature for the same simulations and lines of sight as Fig. \ref{max_direc_deriv_short}, at a frequency of $1.58$ GHz, and we show the corresponding images for a frequency of $0.51$ GHz in Fig. \ref{wav_curv_long}. We find that the polarization wavelength curvature is largest when the derivative of either the polarization intensity or polarization angle with respect to wavelength is close to zero, and that the wavelength curvature tends to be more sensitive to changes in polarization angle, in general. Changes in polarization intensity only appear to be important for supersonic simulations. 

If the magnetic field parallel to the line of sight is small, the shape and sign of the curvature features are similar to those of the rotation measure. If the magnetic field parallel to the line of sight is large, then the curvature is small in regions of high polarization intensity, or high perpendicular component of the magnetic field. There is more small-scale structure for lines of sight parallel to a strong magnetic field, and lines of sight perpendicular to the field have more elongated structures.

We observe that for supersonic simulations there is an increasing amount of small-scale structure as the wavelength increases, and this is also true for lines of sight parallel to a strong field for subsonic simulations. The amplitude of the wavelength curvature tends to increase with wavelength for supersonic simulations with a strong field.

\subsection{A.5 Mixed Derivatives} \label{mixed_deriv}

\begin{figure*}
\begin{center}
\includegraphics[scale=0.58]{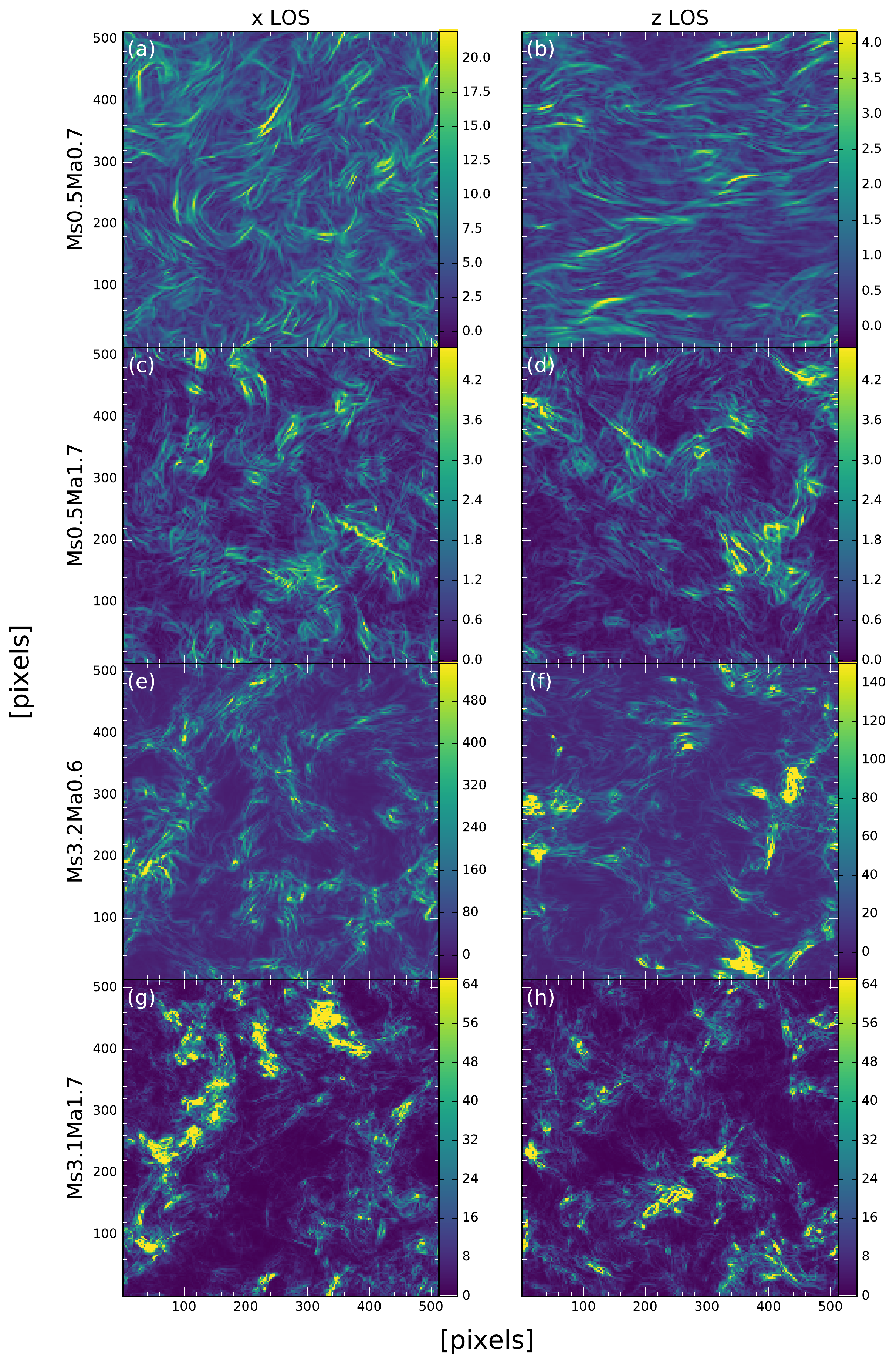}
\caption{The same as Fig. \ref{max_direc_deriv_short}, but for the polarization mixed derivative, instead of the generalized polarization gradient, at a frequency of $1.58$ GHz. Units are pc$^{-1}$ m$^{-2}$. Different color scalings are used for the images.}
\label{max_mix_short}
\end{center}
\end{figure*}

\begin{figure*}
\begin{center}
\includegraphics[scale=0.58]{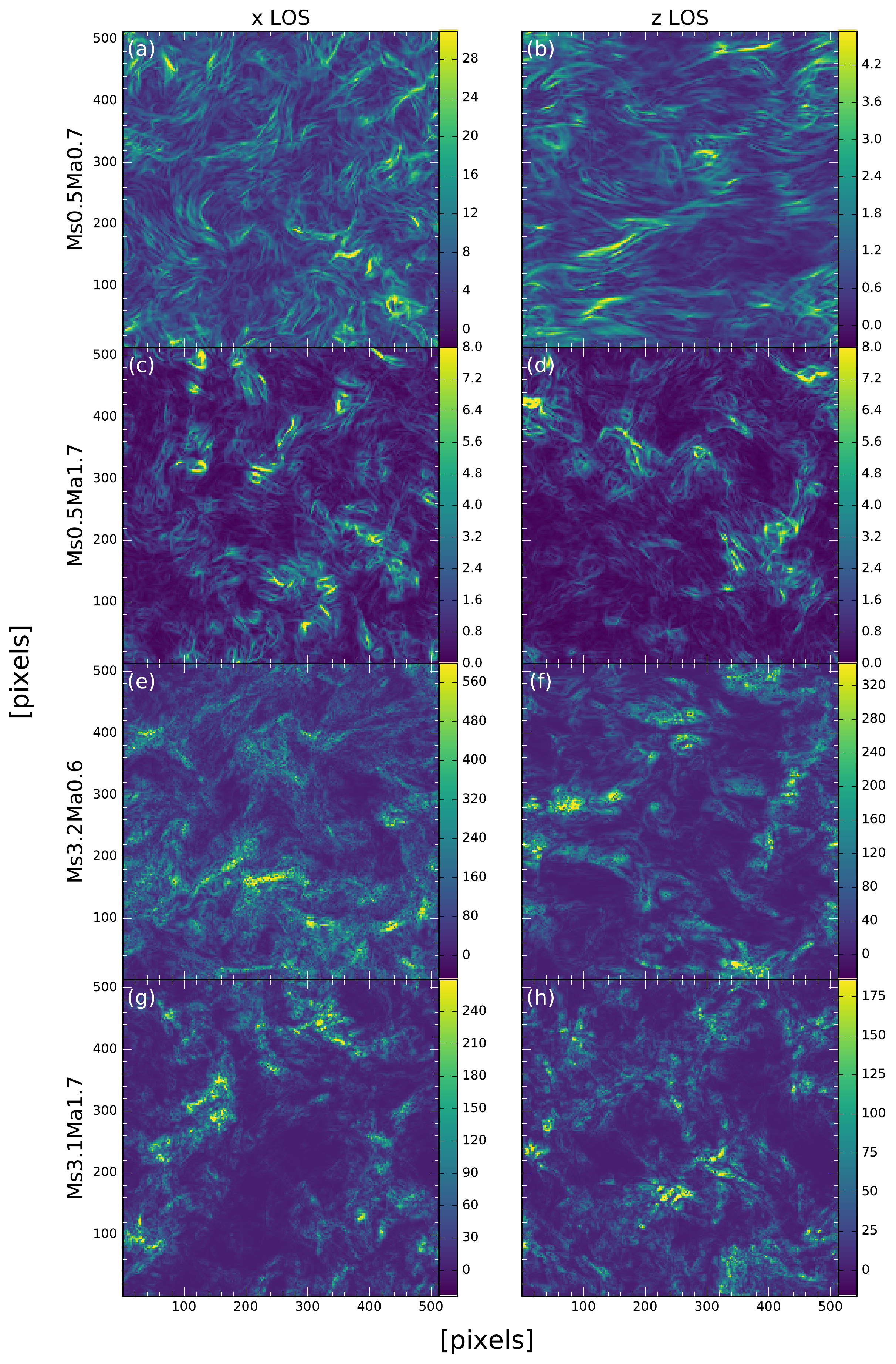}
\caption{The same as Fig. \ref{max_mix_short}, but for a frequency of $0.51$ GHz. Different color scalings are used for the images.}
\label{max_mix_long}
\end{center}
\end{figure*}

We show images of the maximum amplitude of the polarization mixed derivative in Fig. \ref{max_mix_short}, for the same simulations and lines of sight as Fig. \ref{max_direc_deriv_short}, at a frequency of $1.58$ GHz, and at a frequency of $0.51$ GHz in Fig. \ref{max_mix_long}.

For backlit emission, the maximum amplitude of the mixed derivative appears to be equal to the generalized polarization gradient multiplied by the rotation measure. Features are elongated along the field for lines of sight perpendicular to the field, have larger amplitude for lines of sight parallel to a strong field, and tend to be clumped together and less filamentary for supersonic simulations, than for subsonic simulations.

For internal emission, the maximum amplitude of the mixed derivative is similar to the generalized polarization gradient, modulated by the absolute value of the rotation measure. For subsonic simulations, or supersonic simulations with a strong magnetic field parallel to the line of sight, the mixed derivative and generalized polarization gradient become more similar as wavelength increases. For other cases, the mixed derivative and generalized polarization gradient become more different as the wavelength increases. Features tend to be elongated with the field for lines of sight perpendicular to a strong field, and for supersonic simulations, more small-scale structure is apparent as the wavelength increases.

For the angle that maximizes the polarization mixed derivative, lines of sight perpendicular to the field display more elongated structures than lines of sight parallel to the field, and lines of sight parallel to a strong field have more small-scale structure. We find that the angle that maximizes the mixed derivative is correlated with the angles of the gradients of the perpendicular component of the magnetic field and the rotation measure. For subsonic simulations, the degree of correlation increases with wavelength, but for supersonic simulations, the degree of correlation decreases with wavelength.

As for the maximum amplitude of the mixed derivative, lines of sight perpendicular to the field tend to have more elongated features, and lines of sight parallel to a strong field tend to have more small-scale structure, and be more sensitive to wavelength. More small-scale structure appears as the wavelength increases for supersonic simulations, or subsonic simulations and a line of sight parallel to a strong field.

\bibliography{ref_list}
\end{document}